\documentclass[aps,eqsecnum,nofootinbib]{revtex4}
\usepackage[usenames]{color}
\usepackage[dvipsnames]{xcolor}
\usepackage{amssymb, amsmath, amsfonts, amsthm}
\usepackage[utf8]{inputenc} 
\usepackage{float}
\usepackage{flushend}
\usepackage{mathrsfs}
\usepackage{slashed}
\usepackage{bm}
\usepackage{chngcntr}
\usepackage{ytableau}
\usepackage{multirow}

\usepackage{chngcntr}

\counterwithin{table}{subsection} 

\newcommand{\be}{\begin{equation}}
\newcommand{\ee}{\end{equation}}
\newcommand{\bea}{\begin{eqnarray}}
\newcommand{\beas}{\begin{eqnarray*}}
\newcommand{\eea}{\end{eqnarray}}
\newcommand{\eeas}{\end{eqnarray*}}
\newcommand{\ba}{\begin{array}}
\newcommand{\ea}{\end{array}}
\newcommand{\la}{\langle}
\newcommand{\ra}{\rangle}
\usepackage{hyperref}
\usepackage{url}
\hypersetup{
    colorlinks=true, 
    linkcolor=blue,
    filecolor=red,      
    urlcolor=cyan,
    citecolor=magenta}

\begin{document}

\title{CPT Symmetry and Its Breaking in the Chiral Fermion Formalism}

\author{I. P-Castro$^1$}
\email{ivan.perez.c@cinvestav.mx}

\author{J. L. Díaz-Cruz$^2$} 
\email{jldiaz@fcfm.buap.mx}

\author{A. Pérez-Lorenzana$^1$}
\email{aplorenz@fis.cinvestav.mx}

\affiliation{
$^1$Departamento de Física, Centro de Investigación y de 
Estudios Avanzados del I.P.N.\\ Apartado Postal 14-740, 07000, Mexico city, Mexico. \\
$^2$Facultad de Ciencias Físico-Matemáticas, Benemérita Universidad 
Autónoma de Puebla, Apartado Postal 1364, C.P. 72000, Puebla, Pue. México.
}

\begin{abstract}
In this paper, we review the fundamental aspects of ${\cal N} = 1$ supergravity (SUGRA) and employ the chiral formalism to describe spin-$\frac{3}{2}$ fermions. Armed with this, we analyze the discrete symmetries of the graviton in the context of linearized gravity (LG) and Rarita-Schwinger (RS) fields, which include the description of the gravitino. Additionally, we highlight the advantages of the chiral formalism in classifying Lorentz and CPT-violating terms in RS theories, extending these models by including Lorentz-violating operators, as has been done in LG.

\end{abstract}


\maketitle

\section{Introduction}
It is well known that C, P, T, and CP are not always conserved at the quantum level. The Standard Model (SM) incorporates mechanisms to violate these symmetries consistent with experimental observations. However, CPT symmetry is expected to be conserved at any level, as the CPT theorem guarantees. This theorem states that any Lorentz-invariant, local quantum field theory described by a real Lagrangian must conserve CPT symmetry. 

Lorentz symmetry, a cornerstone of modern physics, must be continuously tested experimentally to establish its limits of validity. This has motivated the development of extensions to the SM that include explicit or effective Lorentz violation~\cite{Colladay1997cpt, Colladay1998lorentz}. Importantly, Lorentz violation does not necessarily imply CPT violation, although there are cases where both are linked. In particular, extensions of the SM with Lorentz and CPT violation arise from UV completions in which CPT is spontaneously broken, because a quantum theory of gravity does not admit global symmetries, generating effective field theories with higher-dimensional operators that explicitly break CPT and Lorentz invariance.

Most studies of CPT violation use four-component Dirac fermions, which are convenient for low-energy quantum electrodynamics (QED) since photons provide strong constraints on Lorentz and CPT violation. However, being abundant in the universe, neutrinos can also serve as excellent probes of these symmetries. In the SM and beyond, neutrinos are naturally described by two-component chiral fermions. While Dirac fermions remain a helpful tool, the two-component chiral formalism is particularly advantageous for studying supersymmetric (SUSY) theories, which has become standard practice~\cite{susy2007}. This motivates the formulation of Lorentz-violating extensions of the SM using chiral fields and expressing the full interacting Lagrangian in this framework.

On the other hand, Supergravity (SUGRA) is the unification of general relativity and supersymmetry~\cite{Van1981supergravity, wess2020supersymmetry}. SUSY must manifest as a local symmetry, similar to gauge symmetries in particle physics, which naturally leads to SUGRA as a theory of local supersymmetry. In ${\cal N}=1$ SUGRA in four spacetime dimensions ($d=4$), the gauge multiplet consists of the graviton (spin-$2$ particle) and its supersymmetric partner, the gravitino (spin-$\frac{3}{2}$ particle). 

Investigating local Lorentz symmetry violations in curved spacetime requires a geometric framework that maintains general coordinate invariance while allowing Lorentz invariance breaking. Riemann-Cartan geometry, including torsion alongside curvature, provides a suitable setting for this purpose. Although explicit Lorentz violation in classical gravity theories is inconsistent with geometric principles, spontaneous Lorentz violation is possible~\cite{Kostelecky2004gravity}.

The RS field, which describes particles with spin-$\frac{3}{2}$, plays a key role in both \textit{Chiral Perturbation Theory} (ChPT) and SUGRA. In ChPT, it is used to model baryonic resonances such as the $\Delta(1232)$, improving predictions in low-energy nucleon-pion processes through chiral interaction terms. In SUGRA, the RS field is related under certain considerations to the gravitino, the supersymmetric partner of the graviton, which is essential for maintaining invariance under local supersymmetry. The gravitino is crucial in cosmology and supersymmetry-breaking scenarios, with significant implications for dark matter and quantum gravity studies. For this reason, it is reasonable to incorporate terms in the RS Lagrangian that violate Lorentz invariance.

The organization of this paper is as follows: In Sec.~\ref{Chiral.transformations}, we briefly review chiral field transformations under discrete symmetries. Sec.~\ref{W.gravitational field} discusses massless spin-2 fields and their transformation properties under CP, T, and CPT. Next, In Sec~\ref{Rs. theories} explores RS theories, which describe spin-$\frac{3}{2}$ particles. We show that RS fields can be treated in the chiral formalism by incorporating auxiliary vector fields and discussing their symmetry properties. Next, In Sec.~\ref{Model Extension}, we present the extension of LG and RS theory with terms that violate Lorentz symmetry as follows: In Subsec.~\ref{LIV.LG}, we present an extension of the massless spin-2 theory with spontaneous Lorentz symmetry breaking, showing how CPT violation arises through the breaking of CP and T. Next, in Subsec.~\ref{LIV.RS}, we extends RS theories to include operators that violate CPT, emphasizing the advantages of the chiral fermion formalism. Finally, the Conclusions are summarized in Sec.~\ref{Conclusions}.

\section{Revisiting CP, T, and CPT symmetries in the Chiral fermion formalism} \label{Chiral.transformations}
A non-compact Lie group does not have finite-dimensional unitary irreducible representations. The homogeneous Lorentz group (HLG), $ SO(3,1)$, is non-compact due to the presence of boost generators, $K_i$, and therefore its finite-dimensional representations are not unitary. This can be understood as follows: in the compact case, $SO(4) \cong [SU(2) \times SU(2)] / \mathbb{Z}_2$. By replacing the real parameters in the exponential map associated with $K_i$ by imaginary ones, the group transitions from $SO(4)$ to $SO(3,1)$.
The price we pay is that the resulting representations are no longer unitary, making them unsuitable for implementing Lorentz symmetry on the Fock space of physical states. Then the Lorentz representations can be classified by the representations of the two commuting $SU(2)$ algebras. The irreducible representations of $SU(2)$ are labeled by a single integer or half-integer $j = 0, \frac{1}{2}, 1, \dots$. Therefore, the irreducible representations of the Lorentz group are labeled by two numbers $(j_L, j_R)$. The corresponding representation has dimension $(2j_L + 1)(2j_R + 1)$.

The two-component fermions correspond to two fundamental, non-equivalent irreducible representations of HLG. Proper Lorentz transformations do not mix them but map them within the same functional space. These representations are referred to as chiral representations, typically denoted as ${\cal F} = (\frac{1}{2}, 0)$ and $ \bar{\cal{F}} = (0, \frac{1}{2})$, corresponding to the left- and right-handed components of the fermion field, respectively. Recently, a new approach has been proposed for studying discrete symmetries, where two-component Weyl spinors are treated as the fundamental building blocks. This method, known as the chiral fermion formalism, naturally accommodates CP, T, and CPT symmetries. To illustrate this, let us consider the decomposition of the Dirac field.
    \be
        \Psi(x) = \left(\ba{c} \varphi(x) \\ \bar\chi(x) \ea\right) \, , 
    \ee
To clarify the connection between $ \mathcal{F}$ and $\bar{\mathcal{F}}$, we note that the two-component fermion $\bar{\varphi} = i\sigma^2 \varphi^{*}$ transforms as a fermion in $\bar{\mathcal{F}}$, because $ \sigma^2 \sigma^\mu \sigma^2 = \bar{\sigma}^{\mu T}$, which leads to $\sigma^2 \sigma^{\mu\nu} \sigma^2 = - \sigma^{\mu\nu T}$. Conversely, $ \chi = -i \sigma^2 \bar{\chi}^{*}$ belongs to $\mathcal{F}$. Additionally, dual $\mathcal{F}^{*}$ ($\bar{\mathcal{F}}^{*}$) fermions can be obtained from $\mathcal{F}$ ($\bar{\mathcal{F}}$) fermions using the mapping $\bar{\chi}^\dagger = (i \sigma^2 \chi)^T \, [ \varphi^\dagger = (-i \sigma^2 \bar{\varphi})^T ]$, as their Lorentz transformation properties indicate. Hence, we can express the Lorentz invariants in a convenient two-component notation, where we denote \( \varphi_\alpha \) as fermions of type \( \mathcal{F} \), and \( \bar{\chi}^{\dot{\alpha}} \) as fermions of type \( \bar{\mathcal{F}} \), with \( \alpha = 1,2 \) and \( \dot{\alpha} = \dot{1}, \dot{2} \), respectively.

In this formalism, we can write
    \be
        \bar\chi^\dagger \lambda = \epsilon^{\alpha\beta} \chi_\beta \lambda_\alpha \, , \quad \text{and} \quad \varphi^\dagger \bar\eta = \epsilon_{\dot\alpha\dot\beta} \bar\varphi^{\dot\beta} \bar\eta^{\dot\alpha} \, ,
    \ee
where we define the skew-symmetric matrices as $\epsilon^{\alpha\beta} = (i\sigma^2)^{\alpha\beta}$ and $ \epsilon_{\dot\alpha\dot\beta} = (-i\sigma^2)_{\dot\alpha\dot\beta}$. These definitions follow our previous index conventions. It is also useful to introduce the dual component notation: $\chi^\alpha = \epsilon^{\alpha\beta} \chi_\beta$ for fermions of type $\mathcal{F}^{*}$, and $ \bar\varphi_{\dot\alpha} = \epsilon_{\dot\alpha\dot\beta} \bar\varphi^{\dot\beta}$ for $ \bar{\mathcal{F}}^{*}$-like fermions. The $\epsilon$-matrices serve as tools to lower and raise fermion indices, essentially mapping a fermion to its dual. To ensure consistency, we also have $ \epsilon_{\alpha\beta} = (-i\sigma^2)_{\alpha\beta}$ and $ \epsilon^{\dot\alpha\dot\beta} = (i\sigma^2)^{\dot\alpha\dot\beta}$, such that $\epsilon_{\alpha\beta} \epsilon^{\beta\rho} = \delta_\alpha^\rho$ and $ \epsilon_{\dot\alpha\dot\beta} \epsilon^{\dot\beta\dot\rho} = \delta_{\dot\alpha}^{\dot\rho}$. Following the detailed prescription given in~\cite{IAL2025}. We can establish the discrete transformations:
\begin{itemize}
    \item Under CP, we have
        \begin{equation}
        \ba{c} \varphi_\alpha \,\xrightarrow{~\text{CP}~}\, 
        -i(\sigma^0)_{\alpha\dot\alpha}\bar\varphi^{\dot\alpha}~,
        \\[1em]
        \bar\chi^{\dot\alpha} \,\xrightarrow{~\text{CP}~}\, 
        -i(\sigma^0)^{\dot\alpha\alpha}\chi_{\alpha}~,
        \ea \quad \text{and} \quad
        \ba{c} \bar\varphi_{\dot\alpha} \,\xrightarrow{~\text{CP}~}\, 
        i\varphi^{\alpha}(\sigma^0)_{\alpha\dot\alpha}~,    
        \\[1em]
        \chi^{\alpha} \,\xrightarrow{~\text{CP}~}\,
        i\bar\chi_{\dot\alpha}(\sigma^0)^{\dot\alpha\alpha}~.
        \ea
        \label{cp}
        \end{equation}
    \item Under time reversal, 
        \begin{equation}
        \ba{c} \varphi_\alpha \,\xrightarrow{~\text{T}~}\, 
        -(\sigma^0)_{\alpha\beta}\varphi^{\beta}~,
        \\[1em]
        \bar\chi^{\dot\alpha} \,\xrightarrow{~\text{T}~}\, (\sigma^0)^{\dot\alpha\dot\beta}\bar\chi_{\dot\beta}~,
    \ea
        \quad\text{and}\quad 
        \ba{c} \bar\varphi_{\dot\alpha} \,\xrightarrow{~\text{T}~}\, 
        - \bar\varphi^{\dot\beta}(\sigma^0)_{\dot\beta\dot\alpha}~,
        \\[1em]
        \chi^{\alpha} \,\xrightarrow{~\text{T}~}\,
        \chi_{\beta}(\sigma^0)^{\beta\alpha} ~.
        \ea
        \label{t}
    \end{equation}
    \item Under CPT, 
        \begin{equation}
        \ba{c} \varphi_\alpha \,\xrightarrow{~\text{CPT}~}\, 
        -i{(\sigma^0)_{\alpha}}^{\dot\alpha}\bar\varphi_{\dot\alpha}~,
        \\[1em]
        \bar\chi^{\dot\alpha} \,\xrightarrow{~\text{CPT}~}\, 
        i{(\sigma^0)^{\dot\alpha}}_\alpha\chi^{\alpha}~,
        \ea \quad \text{and} \quad
        \ba{c} \bar\varphi_{\dot\alpha} \,\xrightarrow{~\text{CPT}~}\, 
        i\varphi_{\alpha}{(\sigma^0)^\alpha}_{\dot\alpha}~,
        \\[1em]
        \chi^{\alpha} \,\xrightarrow{~\text{CPT}~}\,
        -i\bar\chi^{\dot\alpha}{(\sigma^0)_{\dot\alpha}}^{\alpha}~.
        \ea
        \label{cpt}
    \end{equation}
\end{itemize}
In this section, we have resumed several conventions regarding phases and transformations within the chiral formalism. However, it is important to note that these conventions are not unique, and alternative conventions can be chosen that still lead to the same results discussed in the subsequent sections and results in~\cite{IAL2025}. For a broader discussion, see Chapter One of \cite{Dreiner2023spinors}, for example. The key distinction between our approach and that in the referenced work lies in our decision to avoid swapping ``representative'' components of chiral fermionic fields in the transformation rules under discrete symmetries. This choice introduces some ambiguities in the transformations under P and C. However, no issues arise with the CP, T, and CPT transformations.

\section{Weak gravitational field: Graviton} \label{W.gravitational field}
The action for the gravitational field is the so-called Einstein-Hilbert (EH) action,
    \be \label{E-H.action}
        S_{\text{EH}} = \frac{1}{2\kappa^2}\int d^4x \sqrt{-g} \, R \, ,
    \ee
where $g := \det (g_{\mu\nu})$, $R$ is the Ricci scalar and $\kappa^2 := 8 \pi G_N$ with $G_N$ the gravitational constant. The vacuum field equations corresponding to the EH action are given by $R_{\mu\nu} = 0$, this action is invariant under local coordinate transformations. However, for our purposes, we will focus on the general properties of the weak gravitational field, which we will call the gravitational field. In this context, the Lagrangian in the EH action should be quadratic in $g_{\mu \nu}$. To obtain this quadratic term, we use the linearized version for~\eqref{E-H.action}, which involves considering small perturbations around the Minkowski metric, {\it Weak field approximation}. Specifically, we introduce a small perturbation $\delta g_{\mu\nu} = \kappa h_{\mu \nu}$ to the metric,
    \be
        g_{\mu \nu} = \eta_{\mu \nu} + \kappa h_{\mu \nu} \, ,
    \ee
where $|\kappa h_{\mu \nu}| \ll 1$, $\eta_{\mu \nu}$ is the Minkowski metric, and $h_{\mu \nu}$ is interpreted as the {\it gravitational field}, the excitations of this field correspond to the graviton, the mediator of gravitational interactions, and has mass dimension.
Expanding the action~\eqref{E-H.action} to linear order in $h_{\mu \nu}$, the Lagrangian density with terms of second-order in $h_{\mu\nu}$ (up to boundary terms) is determined by
    \be \label{E-H action L}
      S_{\text{EH}}^{\text{L}} = \frac{1}{2\kappa} \int d^4x \left( R^{L}_{\mu\nu} - \frac{1}{2} \eta_{\mu\nu} R^L \right) h^{\mu\nu} \, ,  
    \ee
where the linearized Ricci tensor and the linearized Ricci scalar, respectively, are expressed as
    \be \label{T.graviton}
       R^{L}_{\mu\nu} = \frac{1}{2} \kappa \left( \partial^\lambda \partial_\mu h_{\lambda\nu} + \partial^\lambda \partial_\nu h_{\lambda\mu} - \partial^{\lambda}\partial_{\lambda} h_{\mu\nu} - \partial_\mu \partial_\nu h \right) \, , \quad R^L = \eta^{\mu\nu} R^{L}_{\mu\nu} = \kappa (\partial^\mu \partial_\lambda h^\lambda_\mu - \partial^{\lambda}\partial_{\lambda} h )\, ,
    \ee
with $h = \eta^{\mu\nu} h_{\mu\nu}$. Therefore the {\it Linearized Einstein-Hilbert Lagrangian} for the gravitational field is
    \be \label{Linearized Graviton Lagrangian}
        {\cal L}_{\text{EH}}^{\text{L}} = \frac{1}{4} \left[ \partial^\lambda \partial_\mu h_{\lambda\nu} + \partial^\lambda \partial_\nu h_{\lambda\nu} - \partial_\mu \partial_\nu h - \partial^\lambda \partial_\lambda h_{\mu\nu} - \eta_{\mu\nu} \left( \partial^\lambda \partial^\omega h_{\lambda\omega} - \partial^\lambda \partial_\lambda h \right) \right] h^{\mu\nu} \, .
    \ee
A useful way to write this Lagrangian is to notice that we have six terms associated with permutations of indices and using the identity for Lorentzian metric, $\epsilon^{\mu\rho\lambda\delta} \epsilon^{\nu\omega\theta\sigma} \eta_{\delta\sigma} = - 3! {\delta^{[\mu}}_{\nu'} {\delta^{\rho}}_{\omega'} {\delta^{\lambda]}}_{\theta'} \eta^{\nu' \nu} \eta^{\omega\omega'} \eta^{\theta' \theta}$ we can write
    \be \label{EH.L.Lichnerowicz}
        {\cal L}_{\text{EH}}^{L} = -\frac{1}{4} h_{\mu\nu}\epsilon^{\mu\rho\lambda\delta} \epsilon^{\nu\omega\theta\sigma} \eta_{\delta\sigma} \partial_\lambda \partial_\theta h_{\rho\omega} = \frac{1}{4}h_{\mu\nu} \hat{E}^{\mu\nu\rho\omega} h_{\rho\omega} \, .
    \ee
Here $\hat{E}^{\mu\nu\rho\omega}$ is so-called {\it Lichnerowicz operador}.
\subsection{Degrees of freedom of gravitational field: spin content}
The linearized field equations for the gravitational field (wave equation) are
    \be \label{Linearized Einstein equations}
        \partial^\lambda \partial_\mu h_{\lambda\nu} + \partial^\lambda \partial_\nu h_{\lambda\mu} - \partial_\mu \partial_\nu h - \partial^\lambda \partial_\lambda h_{\mu\nu} - \eta_{\mu\nu} \left( \partial^\lambda \partial^\omega h_{\lambda\omega} - \partial^\lambda \partial_\lambda h \right) = 0 \, .
    \ee
To determine the degrees of freedom of the gravitational field, an auxiliary field is usually used,$\bar{h}_{\mu\nu} = h_{\mu\nu} - \frac{1}{2} \eta_{\mu\nu} h$, with which the equations of motion are reduced to $\bar{R}_{\mu\nu}^{\text{L}} = 0$ and with the so-called de Donder gauge, $\partial^\mu \bar{h}_{\mu\nu} = 0$. However, this does not completely fix the gauge because 
    \be
        \delta \left( \partial_\rho h^{\rho\nu} - \frac{1}{2} \partial^\nu {h^\rho}_\rho \right) = \Box \xi^\nu \, .
    \ee
Instead, a better choice for counting degrees of freedom (in $d \geq 4$) is to choose a gauge analogous to the Coulomb gauge, given by $d$-conditions
    \be \label{EH.L.gauge}
        \partial^ih_{i\mu} = 0 \, ,
    \ee
where the sum includes only the spatial coordinates $i = 1, \ldots, d-1$. In this scheme it can be shown that $\xi^\mu = 0$. Analyzing the components $\mu, \nu$: $\{0,0\}, \{i,i\}, \{0,i\}$ from the wave equation given in \eqref{Linearized Einstein equations} we obtain, respectively
    \be
        \nabla^2 h_{ii} = 0 \quad , \quad -2\partial_0^2 h_{ii} + \nabla^2 h_{ii} - \nabla^2 h_{00} = 0 \quad \text{and} \quad \nabla^2 h_{0i} = 0 \, .
    \ee
With this, it is possible to show that the only non-zero components are of the form $h_{ij}$ and $h = 0$, plus the gauge condition $\partial^i h_{ij} = 0 $ we can fix other $(d-1)$ degrees of freedom, with the condition of null trace we obtain finally $ \frac{d(d-1)}{2} - (d-1) -1 = \frac{d(d-3)}{2}$ degrees of freedom. For $d = 4$, two possible polarizations of the graviton are taken into account by introducing the polarization tensor $h_{\mu\nu}^{(\lambda_1 \lambda_2)}$. The latter can be composed of the usual polarization vectors (for spin-$1$). For example, these vectors could satisfy the relations
$\eta^{\mu\nu}\epsilon_\mu^{(\lambda) *}\epsilon_\nu^{(\lambda)} = -1 , \eta^{\mu\nu}\epsilon_\mu^{(\lambda)} \epsilon_\nu^{(\lambda)} = 0$
and obey the Lorentz gauge $k^\mu \epsilon_\mu = 0$. We can form the polarization tensor $h_{\mu\nu}^{(\lambda_1\lambda_2)} := \epsilon_\mu^{(\lambda_1)} \epsilon_\nu^{(\lambda_2)}$. And the plane wave decomposition of $h_{\mu\nu}$ is then written as
    \be \label{h.quantum}
        h_{\mu\nu} (x) = \sum_{\lambda_1 \lambda_2}\int \widetilde{d^3p} \left[ a(\vec{p}, \lambda_1\lambda_2) h_{\mu\nu}^{(\lambda_1\lambda_2)}(\vec{p}) e^{-i p\cdot x} + a^\dagger(\vec{p},\lambda_1 \lambda_2) h_{\mu\nu}^{(\lambda_1\lambda_2)*}(\vec{p})e^{i p\cdot x}\right] \, ,
    \ee
with $\widetilde{d^3p} = d^3p/2E(2\pi)^3$ and the sum is over all graviton helicities $\lambda_1 \lambda_2 = ++, --$.
    \be
    \begin{aligned}
        \text{helicity}= 2: \quad h_{\mu\nu}^{(++)} &= \epsilon_\mu^{(+)} \epsilon_\nu^{(+)} \, , \\
        \text{helicity}= -2: \quad h_{\mu\nu}^{(--)} &= \epsilon_\mu^{(-)} \epsilon_\nu^{(-)} \, .
    \end{aligned} \label{tensor.pol}
    \ee
As is well known, when studying massive particles with spin greater than $\frac{1}{2}$, it is not possible to construct an equivalent representation to the massless case with the creation and annihilation operators that describe helicity $\pm j$ for $j\geq 1$ (see Ch.~5 of \cite{Weinberg1995quantum}). We can understand this as follows: for massive particles, polarization vectors certainly exist, but they must satisfy certain constraints, such as the Lorentz gauge. These constraints arise because we use fields as auxiliary objects to describe the interactions of more fundamental particles. For massless particles in four dimensions, if we choose polarization vectors $\epsilon^{(\lambda) \mu}$ for photons of helicity $\pm 1$, we find that under Lorentz transformations, the momentum transforms as $(\Lambda k)^\mu = k^\mu$, while the polarization vector transforms as
    \be
        (\Lambda \epsilon^{(\lambda)})^\mu = e^{i \theta} \epsilon^{(\lambda)\mu} + \beta(\Lambda, k) k^\mu.
    \ee
This shows that polarization vectors do not transform as true vectors. Instead, only the ``gauge equivalence class'' $\{\epsilon^{(\lambda)\mu} \,| \, \epsilon^{(\lambda)\mu} + \beta k^\mu\}$ is invariant under Lorentz transformations. Therefore, we do not have simple transformations under the HLG. In fact, who has simple transformations is $F_{\mu\nu} = \partial_\mu A_\nu - \partial_\nu A_\mu$ which transforms like $(1,0) \oplus (0,1)$. And here $h_{\mu\nu}$ must transform as $(1,1)$ representation under the HLG. If we employ the prescription dad in Eq.~\eqref{tensor.pol} it is clear that the quantization for the field $h_{\mu\nu}$ is inconsistent as is the quantization of $A^{\mu}$, but its associated polarization vectors are useful in the calculation of amplitudes.

With the discovery of gravitational waves, it is reasonable to consider that if gravitons exist, they could have zero mass. In the case that interests us, we can construct an object suitable for the description of the graviton as follows: we can build a rank $4$ tensor (${R^{\text{L} \lambda}}_{\mu\rho\nu}$, linearized Riemman tensor) that, in principle, transforms as $[(1,0) \oplus (0,1)] \otimes [(1,0) \oplus (0,1)]$, this representation has 36 degrees of freedom. However, using antisymmetry between a pair of indices and symmetry between pairs of indices, it can be shown that only $21$ elements are independent, while the Bianchi identity provides an additional constraint, leaving $20$ independent elements. The contraction of this tensor, $R_{\mu\nu}^{\text{L}} := {R^{\text{L} \lambda}}_{\mu\lambda\nu} = {\eta_\lambda}^\rho {R^{\text{L} \lambda}}_{\mu\rho\nu}$, behaves like the so-called Linearized Ricci tensor with $10$ independent elements, and as we have seen in this section, we can express $R_{\mu\nu}^{\text{L}}$ in terms of $h_{\mu\nu}$ (see Eq.~\eqref{T.graviton}). Using \eqref{h.quantum}, it is evident that one can construct the representation $R_{\mu\nu}^{\text{L}} \equiv (2,0) \oplus (0,2)$ with the annihilation and creation operators for massless particles of spin $2$. The presence of the derivatives in the first Eq.~\eqref{T.graviton} indicates that an interaction density built solely from $R_{\mu\nu}^{\text{L}}$ and its derivatives will have matrix elements that vanish more rapidly for small massless particle energies and momenta than one using the tensor field $h_{\mu\nu}$. Interactions in such a theory will have a correspondingly rapid decay at large distances, faster than the usual inverse-square law. Therefore, to incorporate the usual inverse-square gravitational interactions, it is best to work with $h_{\mu\nu}$, and as we shall see below, its transformation laws can be adequately defined under discrete symmetry transformations, i.e., under total Lorentz Group $O(1,3)$.
\subsection{Gravitational field under discrete symmetries} \label{LG.Transformations}
LG is a local theory because its equations of motion and fundamental principles adhere to the principle of locality. This means that interactions occur at specific points in the space-time, without any instantaneous action at a distance. The field $h_{\mu\nu}$ under proper Lorentz transformations transforms according to its Lorentz indices, 
    \be \label{L.Transformation LG}
        h_{\mu\nu}(x) \rightarrow h_{\mu\nu}^\prime(x) = {\Lambda_\mu}^\rho {\Lambda_\nu}^\omega h_{\rho\omega} (\Lambda^{-1}x)
    \ee
We will have a Lorentz invariant local theory, so it must be CPT invariant.

We are interested in finding out whether or not graviton Lagrangian changes under discrete symmetries, we would expect it to remain invariant. The line element in a spacetime described by $g_{\mu\nu}$ is $ds^2 = g_{\mu\nu} d x^{\mu} dx^{\nu}$. Since $g_{\mu\nu} =  \eta_{\mu\nu} + \kappa h_{\mu\nu}$, we can expect $g_{\mu\nu}$ to inherit some properties of $\eta_{\mu\nu}$ like $\eta^{\mu\lambda}\eta_{\lambda\nu} = {\delta^\mu}_{\nu}$. In the chiral formalism, we can express the Minkowski metric tensor with in the following way
    \be
        \eta_{\mu\nu} = \frac{1}{2}\epsilon^{\alpha\beta} \epsilon^{\dot\alpha\dot\beta} (\sigma_\mu)_{\alpha\dot\alpha} (\sigma_\nu)_{\beta\dot\beta} \, .
    \ee
For this reason it is direct to see that $({\cal CP})^{-1} \eta_{\mu\nu} {\cal CP} = \eta_{\mu\nu}$ since the Infeld–Van der Waerden symbols do not change under CP and since $\epsilon^{\alpha\beta} = (i\sigma^2)^{\alpha\beta}$ ($\epsilon^{\dot\alpha\dot\beta} = (i\sigma^2)^{\dot\alpha\dot\beta}$). Under T, We must take into account that ${\cal T}^{-1} \epsilon^{\alpha\beta} {\cal T} = \epsilon^{\alpha\beta}$ (${\cal T}^{-1} \epsilon^{\dot\alpha\dot\beta} {\cal T} = \epsilon^{\alpha\beta}$) such that ${\cal T}^{-1} \sigma^2 {\cal T} = \sigma^{2 *} = -\sigma^2$ and ${\cal T}^{-1} i {\cal T} = -i$. Furthermore, we have to 
    \be
        \eta_{\mu\nu} = \, \xrightarrow{~~\text{T}~~} \, \frac{1}{2} \epsilon^{\beta\alpha} \epsilon^{\dot\beta\dot\alpha} (\sigma_\mu^{*})_{\alpha\dot\alpha} (\sigma_\nu^{*})_{\beta\dot\beta} = {T_\mu}^\lambda {T_\nu}^\omega \eta_{\lambda\omega} 
        = \eta_{\mu\nu} \, .
    \ee
We have used $\sigma_\mu^{*} = -{T_\mu}^\nu \bar\sigma_\nu^{T}$. On the other hand, the classical description of gravity in a Riemannian space-time uses a metric and a covariant derivative acting on vector representations or tensors of the group $ \textrm{Gl}(4, \mathbb{R}) $. However, $ \textrm{Gl}(4, \mathbb{R}) $ has no spinor representations, whereas the fundamental constituents of ordinary matter, such as leptons and quarks, are known to be excitations of spinor fields. For this reason, it is not possible to obtain an expression for $h_{\mu\nu}$ in terms of only chiral indices. Now, we can set the condition that the gravitational field, $h_{\mu\nu}$ must satisfy to keep CP, T, and CPT. We rewrite the graviton Lagrangian for convenience as (see Eq.~\eqref{EH.L.Lichnerowicz})
    \be \label{Graviton Lagrangian}
        {\cal L}_{\text{graviton}} = \frac{1}{4} h_{\mu\nu} \epsilon^{\mu\rho\lambda\delta} \epsilon^{\nu\omega\theta\sigma} \eta_{\delta\sigma} i\partial_\lambda i\partial_\theta h_{\rho\omega}  \, .
    \ee
Under CP: $i\partial_{\mu}  \,\xrightarrow{~\text{CP}~}\, {P_\mu}^\nu i\partial_\nu$ and $ \epsilon^{\mu\rho\lambda\delta}  \,\xrightarrow{~\text{CP}~}\, -\epsilon^{\mu\rho\lambda\delta}$, While under T: $i\partial_{\mu}  \,\xrightarrow{~\text{T}~}\, -{T_\mu}^\nu i\partial_\nu, \epsilon^{\mu\rho\lambda\delta}  \,\xrightarrow{~\text{T}~}\, -\epsilon^{\mu\rho\lambda\delta}$, then
    \be
        {\cal L}_{\text{graviton}} \xrightarrow{~\text{CP}~} {P_\lambda}^\varsigma {P_\theta}^\vartheta \frac{1}{4} h_{\mu\nu}^{\text{CP}} \epsilon^{\mu\rho\lambda\delta} \epsilon^{\nu\omega\theta\sigma} \eta_{\delta\sigma} i\partial_\varsigma i\partial_\vartheta h_{\rho\omega}^{\text{CP}} \quad \text{and} \quad {\cal L}_{\text{graviton}} \xrightarrow{~\text{T}~} {T_\lambda}^\varsigma {T_\theta}^\vartheta h_{\mu\nu}^{\text{T}} \frac{1}{4} \epsilon^{\mu\rho\lambda\delta} \epsilon^{\nu\omega\theta\sigma} \eta_{\delta\sigma} i\partial_\varsigma i\partial_\vartheta h_{\rho\omega}^{\text{T}}  .
    \ee
We could propose.
    \be
        h_{\mu\nu} \xrightarrow{~~\text{CP}~~} h^{\text{CP}}_{\mu\nu} = {P_\mu}^\rho {P^\tau}_\nu h_{\rho\tau} \quad \text{and} \quad h^{\mu\nu} \xrightarrow{~~\text{CP}~~} {h^{\text{CP}}}^{\mu\nu} = {P^\mu}_\rho {P_\tau}^\nu h^{\rho\tau}\, .
    \ee
Note that in principle, we could add some phases due to charge conjugation; however, gravitons have no charge to change, so that phase was taken with a factor of 1. With these transformations, \eqref{Graviton Lagrangian} remains invariant under CP. And 
    \be
        h_{\mu\nu} \xrightarrow{~~\text{T}~~} h^{\text{T}}_{\mu\nu} = {T_\mu}^\rho {T^\tau}_\nu h_{\rho\tau} \quad \text{and} \quad h^{\mu\nu} \xrightarrow{~~\text{T}~~} {h^{\text{T}}}^{\mu\nu} = {T^\mu}_\rho {T_\tau}^\nu h^{\rho\tau}\, ,
    \ee
since ${T_\mu}^\rho {T_\rho}^\nu = {\delta_\mu}^\nu$. It is evident that the Lagrangian for graviton is time reversal invariant.

With the previous transformations, it is straightforward to observe that
    \be
        h_{\mu\nu} \, \xrightarrow{~~\text{CPT}~~} \, h_{\mu\nu} \quad \text{and} \quad
        h^{\mu\nu} \, \xrightarrow{~~\text{CPT}~~} \, h^{\mu\nu}\, .
    \ee
Given that $i\partial_{\mu}  \,\xrightarrow{~\text{CPT}~}\, i\partial_\mu$, it is clear that ${\cal L}_{\text{graviton}}$ is CPT invariant, according to the CPT theorem. Note that the proposed transformations for CP implicitly clarify the transformations under P and C. However, we will focus on other transformations in order to work within the framework where the chiral formalism is unambiguous.
\section{Rarita-Schwinger field} \label{Rs. theories}
The main problems in local high-spin field theory are closely related to the presence of unphysical low-spin components in the covariant representation of the field. For example, the RS field exhibits unphysical spin-$1/2$ components in addition to its spin-$3/2$ components. However, in SUGRA, the incorporation of the free field can be done successfully.

In this subsection, we will consider the corresponding transformations of the RS field under discrete symmetries (CP, T, and CPT). The general RS Lagrangian has the following form (see e.g.~\cite{Moldauer1956properties, Nath1971uniqueness})
    \be \label{RS.Field.general.1}
        {\cal L}_{\text{RS}}^{\text{general}} = \bar{\Psi}_\mu \Lambda^{\mu\nu} \Psi_\nu \, ,
    \ee
here the field $\Psi^\mu$ is the RS field. The most general $\Lambda^{\mu\nu}$ subject to the following four conditions:
    \begin{enumerate}
    \item The Euler-Lagrange equations derived from the free action~\eqref{RS.Field.general.1} should yield the local Rarita-Schwinger equations for a spin-$\frac{3}{2}$ particle. These equations consist of a Dirac equation for each vector-spinor component, as well as supplementary conditions to remove the lower spin degrees of freedom:
        \be \label{Conditions}
            (\slashed{p} - m) \Psi_\mu = 0, \quad \text{and} \quad \gamma^\mu\Psi_\mu = 0 \, .
        \ee
    \item The Lagrangian should be non-singular in the limit $p \rightarrow 0$. In particular, we would like the pole of the propagator to occur at the particle's mass.
    \item The Lagrangian should be linear in derivatives because it describes a fermionic field.
    \item The Lagrangian must be Hermitian, equivalently, the operator $\Lambda^{\mu\nu}$ must satisfy the hermiticity condition
        \be \label{Gen.cond}
            \gamma^0 (\Lambda^{\mu\nu})^\dagger \gamma^0 =
                \Lambda^{\nu\mu}
        \ee
\end{enumerate}
has the following form~\cite{Moldauer1956properties,Nath1971uniqueness}
    \be \label{RS.Op}
        \Lambda^{\mu\nu} = (\not{p} - m) \eta^{\mu\nu} + A (\gamma^\mu p^\nu + \gamma^\nu p^\mu) + \frac{1}{2} B(A) \gamma^\mu \not{p} \gamma^\nu + m C(A) \gamma^\mu \gamma^\nu \, ,
    \ee
where $B(A) = 3A^2 + 2A + 1$ and $C(A) = 3A^2 + 3A +1$ with $A$ a real parameter, but $A \neq \frac{1}{2}$. The Lagrangian Eq.~\eqref{RS.Field.general.1} is invariant under the contact/point transformation 
    \be
        \Psi^\mu \rightarrow {\Psi^\prime}^\mu = (\eta^{\mu\nu} + \alpha \gamma^\mu \gamma^\nu) \Psi_\nu = \Theta^{\mu\nu}(\alpha) \Psi_\nu , \quad A \rightarrow A^\prime = \frac{A - 2\alpha}{1 + 4\alpha} \, ,
    \ee
with parameter $\alpha \neq -\frac{1}{4}$. We can rewrite $\Lambda_{\mu\nu}(A)$
as follow
    \be 
        \Lambda_{\mu\nu}(A) = \Theta_{\mu\alpha}(\alpha) \Lambda^{\alpha\beta}(A^\prime) \Theta_{\beta\nu}(\alpha) \, .
    \ee
Here $\alpha$ is real. Note if we parameterize $\alpha \to \alpha = \frac{a-1}{4}$, it is easy to show that for $a \in \mathbb{R}$, the contact/point transformations with $a \neq 0$ (equivalent to $\alpha \neq - \frac{1}{4}$) form a group that is isomorphic to the group formed by $\mathbb{R} \backslash \{0\}$ with the usual operation of multiplication.

A consequence of equation~\eqref{Conditions} is the constrain $p^\mu \Psi_\mu = 0$. We can rewrite ~\eqref{RS.Field.general.1}, using the identities $\gamma^\mu \gamma^\nu = \eta^{\mu\nu} - i \Sigma^{\mu\nu}$ and $\gamma^\mu\gamma^\lambda\gamma^\nu = \eta^{\mu\lambda} \gamma^\nu + \eta^{\lambda\nu} \gamma^\mu - \eta^{\mu\nu} \gamma^\lambda - i\epsilon^{\rho\mu\lambda\nu} \gamma_\rho \gamma_5$.
    \bea \label{Rarita-Schwigner Lagrangian}
        {\cal L}_{\text{RS}}^{\text{general}} &&= \left( 1 - \frac{1}{2}B(A) \right) i \eta^{\mu\nu} \bar{\Psi}_\mu \gamma^{\lambda} \partial_\lambda \Psi_\nu + \left( \frac{B(A)}{2} + A \right) i \eta^{\mu\lambda} \bar{\Psi}_\mu \gamma^\nu \partial_\lambda \Psi_\nu + \left( \frac{B(A)}{2} + A \right) i \eta^{\lambda \nu} \bar{\Psi}_\mu \gamma^\mu \partial_\lambda \Psi_\nu \nonumber \\
        &&\quad + \frac{1}{2} B(A) \epsilon^{\mu\rho\lambda\nu} \bar{\Psi}_\mu \gamma_5 \gamma_\rho \partial_\lambda \Psi_\nu + (C(A)-1) m \eta^{\mu\nu} \bar{\Psi}_\mu \Psi_\nu - iC(A)m \bar{\Psi}_\mu \Sigma^{\mu\nu} \Psi_\nu \, ,
    \eea
here $\Sigma^{\mu\nu} = \frac{i}{2} [\gamma^\mu, \gamma^\nu]$ and $\epsilon^{\mu\lambda\rho\nu}$ (with $\epsilon^{0123}=+1$) is Levi-Civita symbol.
\subsection{Degrees of freedom of massive Rarita-Schwinger field: spin content}
The RS field is a spinor-vector, so it must transform under the Lorentz group as the product representation: $(\frac{1}{2}, \frac{1}{2}) \otimes [(\frac{1}{2},0) \oplus (0,\frac{1}{2})]$, i.e. four-vector combined with spinor of defined chirality. Then, we can write the RS field in terms of chiral vector-spinors, i.e.
    \be \label{RSfield}
        {\Psi}^{\mu} (x) =\left( \ba{c} {\varphi}^{\mu}(x) \\ {\bar\chi}^{\mu}(x) \ea\right) ~, \quad  \mu=0,1,2,3 ~.
    \ee
The vector-spinor field thus contains two spin-$\frac{1}{2}$ components in addition to the physical spin-$\frac{3}{2}$ component. One may identify $p^\mu \Psi_\mu$ with $\left( \frac{1}{2}, 0\right) \oplus \left ( 0, \frac{1}{2}\right)$ once it is realized that $p^\mu \Psi_\mu$ transforms in the same way as a Dirac field i.e.,
    \be \label{RS.Rep.1}
        \left(\frac{1}{2} , 0\right) \equiv p^\mu \varphi_\mu \quad \text{\and} \quad \left( 0, \frac{1}{2} \right) \equiv p^\mu \bar\chi_\mu \, .
    \ee
The remaining part of the $\Psi^\mu$ which is complementary to~\eqref{RS.Rep.1} is then given by
    \be \label{Rs.Compl.1}
        \left( 1, \frac{1}{2} \right) \oplus \left( \frac{1}{2} , 1 \right) \equiv \left( \eta^{\mu \nu} - \frac{p^\mu p^\nu}{p^2} \right) \Psi_\nu \, ,
    \ee
which has a zero contraction with $p_\mu$.
Eq.~\eqref{Rs.Compl.1} contains spin-$\frac{3}{2}$ proper as well as an auxiliary spin-$\frac{1}{2}$ contribution.

For which using the ``Gram-Schmidt orthogonalization method'', with the vectors $p^\mu, \gamma^\mu$ and $\Psi^\mu$. Heuristically we can find the following spin projectors (see App.~\ref{D.red}) in $d=4$:
    \be \label{General spin projectors.4}
    \begin{aligned}
        (P^{3/2})^{\mu\nu} &= \frac{1}{3p^2} \left[ 3p^2 \eta^{\mu\nu} - \gamma^\mu \gamma^\nu p^2 - \slashed{p} \left( \gamma^\mu p^\nu - p^\mu \gamma^\nu \right) - 2 p^\mu p^\nu \right] \\
        (P^{1/2})^{\mu\nu} &= \frac{1}{3p^2} \left[ \gamma^\mu \gamma^\nu p^2 + \slashed{p} \left( \gamma^\mu p^\nu - p^\mu \gamma^\nu \right) + 2 p^\mu p^\nu \right] = (P^{1/2}_{11})^{\mu\nu} + (P^{1/2}_{22})^{\mu\nu}  \\
        \left( P^{1/2}_{11} \right)^{\mu \nu} &= \frac{1}{3 p^2} \left( \gamma^\mu \gamma^\nu p^2 + \slashed{p} \left( \gamma^\mu p^\nu - p^\mu \gamma^\nu \right) - p^\mu p^\nu \right) \\
        \left( P^{1/2}_{22} \right)^{\mu \nu} &= \frac{p^\mu p^\nu}{p^2} \\
        \left( P^{1/2}_{12} \right)^{\mu \nu} &= \frac{1}{\sqrt{3}p^2} \left(p^\mu p^\nu - \slashed{p}\gamma^\mu p^\nu \right) \\
        \left( P^{1/2}_{21} \right)^{\mu \nu} &= \frac{1}{\sqrt{3}p^2} \left(\slashed{p} p^\mu \gamma^\nu  - p^\mu p^\nu \right)  \, .
    \end{aligned}
    \ee
From the condition  $\gamma^\mu \Psi_\mu = 0$ we can obtain the condition $(\sigma^\mu)_{\alpha \dot\beta} (\bar\chi_\mu)^{\dot\beta} = 0$ ($(\bar\sigma^\mu)^{\dot\alpha\beta} (\varphi_\mu)_\beta = 0$), where we now explicitly write the spinor indices, represents a constraint equation for each value of the spin index $\alpha$ ($\dot\alpha$), While the condition $\partial_\mu \Psi^\mu = 0$ leads us to the condition $\partial_\mu (\varphi^\mu)_\beta = 0$ ($\partial_\mu (\bar\chi^\mu)^{\dot\beta} = 0$) is an equation of motion for the spinor components $(\varphi^0)_\beta$ ($(\bar\chi^0)^{\dot\beta}$). This is
    \be
        \partial_0 (\bar\chi^0)^{\dot\beta} = - \partial_j (\bar\chi^j)^{\dot\beta}  \quad \text{with} \quad (\bar\chi^0)^{\dot\beta} = - {( \sigma^j)^{\dot\beta}}_{\dot\alpha} (\bar\chi_j)^{\dot\alpha} \, ,
    \ee
    \be
        \partial_0 (\varphi^0)_\beta = - \partial_j (\varphi^j)_\beta \quad \text{with} \quad (\varphi^0)_\beta = - {(\bar\sigma^j)_\beta}^{\alpha}(\varphi_j)_\alpha \, .
    \ee
However, the Dirac-like equation~\eqref{Conditions} also gives an equation of motion for the same vector-spinor components,
    \be \label{Motion equations.G}
        i (\sigma^\nu)_{\alpha\dot\beta} \partial_\nu (\bar\chi_\mu)^{\dot\beta} = m (\varphi_\mu)_\alpha \quad \text{and} \quad i(\sigma^\nu)^{\dot\alpha\beta} \partial_\nu (\varphi_\mu)_\beta = m (\bar\chi_\mu)^{\dot\alpha} \, .
    \ee
and when taken together, these result in another set of constraints. These two sets of equations each constitute four constraints and serve to remove $8$ components of the $16$-component vector-spinor $\Psi^\mu$, leaving $2(2s+1) = 8$ physical degrees of freedom as required for a massive spin $s = \frac{3}{2}$ particle. In conclusion, the vector-spinor $\Psi^\mu$ has $16$ particle states distributed as (we have two spin-$\frac{1}{2}$ different sectors):
    \be
        \text{spin} \, \frac{3}{2} = \begin{cases} 
        4 \quad \text{particles} \\
        4 \quad \text{anti-particles}
        \end{cases} \, , \quad \text{spin} \, \frac{1}{2} = \begin{cases} 
        2 \quad \text{particles} \\
        2 \quad \text{anti-particles}
        \end{cases} \, , \quad \text{spin} \, \frac{1}{2} = \begin{cases} 
        2 \quad \text{particles} \\
        2 \quad \text{anti-particles}
        \end{cases}
    \ee
To describe a massive  particle of spin-$3/2$, we require
    \be
        \Psi_{3/2}^\mu = (P^{3/2})^{\mu\nu} \Psi_\nu \, .
    \ee
From the general RS Lagrangian, we can choose the particular case in which $A=-1$, then this reduces to
    \be \label{Massive gravitino}
        {\cal L}_{\rm RS}(A =-1) = \bar\Psi_\mu \left( \epsilon^{\mu\rho\lambda\nu} \gamma_5 \gamma_\rho \partial_\lambda - im \Sigma^{\mu\nu} \right) \Psi_\nu = 2i {\cal L}_{\rm gravitino}\, .
    \ee
This formulation is used to describe a massive gravitino (see Chapter 31 of ~\cite{Weinberg2005quantum}).
\subsection{Rarita-Schwinger field under discrete symmetries}
Notice that the derivative in the kinetic term does not affect the field transformations. Using the usual Dirac covariant bilinears, we can determine analogous covariant bilinear currents with the RS field,
    \be
        \ba{rclcrcl}
        \bar{\Psi}_\mu \gamma^\lambda \Psi_\nu &=& \chi_\mu \sigma^\lambda \bar{\chi}_\nu + \bar{\varphi_\mu} \bar\sigma^\lambda \varphi_\nu &\qquad&
        \bar\Psi_{\mu} \gamma_\lambda\gamma_5\Psi_{\nu} &=& \chi_\mu \sigma_\lambda \bar\chi_\nu - \bar\varphi_\mu \bar\sigma_\lambda \varphi_\nu  \nonumber\\
        \bar{\Psi}_\mu \gamma^\nu \Psi_\nu &=& \chi_\mu \sigma^\nu \bar{\chi}_\nu + \bar{\varphi_\mu} \bar\sigma^\nu \varphi_\nu &\quad&
        \bar{\Psi}_\mu \Psi_\nu &=& \bar\varphi_\mu \bar\chi_\nu + \chi_\mu \varphi_\nu \nonumber\\
        \bar{\Psi}_\mu \gamma^\mu \Psi_\nu &=& \chi_\mu \sigma^\mu \bar{\chi}_\nu + \bar{\varphi_\mu} \bar\sigma^\mu \varphi_\nu && \bar{\Psi}_\mu \Sigma^{\mu\nu} \Psi_\nu &=& 2i \chi_\mu \sigma^{\mu\nu} \varphi_\nu + 2i \bar\varphi_\mu \bar\sigma^{\mu\nu} \bar\chi_\nu \, .
        \ea
        \label{Dcurr.RS}
    \ee
Thus, we can write the RS Lagrangian \eqref{Rarita-Schwigner Lagrangian} in the chiral formalism as
    \be
    \begin{aligned} \label{RS Lagrangian.Chiral}
        {\cal L}_{\text{RS}}^{\text{general}} &= \left( 1 - \frac{1}{2}B(A) \right) \eta^{\mu\nu} (\chi_\mu \sigma^\lambda i\partial_\lambda \bar{\chi}_\nu + \bar{\varphi_\mu} \bar\sigma^\lambda i\partial_\lambda \varphi_\nu) + \left( \frac{B(A)}{2} + A \right) \eta^{\mu\lambda} ( \chi_\mu \sigma^\nu i\partial_\lambda \bar{\chi}_\nu + \bar{\varphi_\mu} \bar\sigma^\nu i\partial_\lambda \varphi_\nu ) \\
        &\quad + \left( \frac{B(A)}{2} + A \right) \eta^{\lambda\nu} (\chi_\mu \sigma^\mu i\partial_\lambda \bar{\chi}_\nu + \bar{\varphi_\mu} \bar\sigma^\mu i\partial_\lambda \varphi_\nu) + \frac{i}{2} B(A) \epsilon^{\mu\rho\lambda\nu} (\chi_\mu \sigma_\lambda i\partial_\rho \bar\chi_\nu - \bar\varphi_\mu \bar\sigma_\lambda i\partial_\rho \varphi_\nu) \\
        &\quad + (C(A)-1) m \eta^{\mu\nu} (\bar\varphi_\mu \bar\chi_\nu + \chi_\mu \varphi_\nu) + 2C(A) m ( \chi_\mu \sigma^{\mu\nu} \varphi_\nu + \bar\varphi_\mu \bar\sigma^{\mu\nu} \bar\chi_\nu) \, .
    \end{aligned}
    \ee
As we have seen, chiral formalism is a natural framework for describing CP, T, and CPT. Therefore, in the case studied, we will see if the general RS Lagrangian is invariant under these symmetries. To discuss what happens to the Lagrangian under certain transformations, we must indicate how the four-vector $B_\mu$ changes under those transformations. It is important to note that this corresponds directly to a Lorentz vector such as $A_\mu$ and $\partial_\mu$, which are associated with the representation $(\frac{1}{2}, \frac{1}{2})$. Then, we can associate changes related to parity, charge conjugation, and time reversal (see, for example, ~\cite{IAL2025}).

Under parity, we simply have ${\cal P}^{-1} B_{\mu}{\cal P} ={P_\mu}^{\nu}B_{\nu}$, analogously under time reversal, we have ${\cal T}^{-1} B_\mu{\cal P} = {T_\mu}^\nu B_\nu$. The question now is to indicate how they would change under charge conjugation, in general a phase would appear, say ${\cal C}^{-1} B_{\mu} {\cal C} = e^{i\phi_{B}}B_{\mu}$. For simplicity, we will assume that we have two possibilities: that it changes or does not change with a phase equivalent to a change of sign, just as happens with the electromagnetic 4-potential. Then ${\cal C}^{-1} B_\mu{\cal C} = \eta_{B}B_\mu$, such that $\eta_B = +1$ if there is no change and $\eta_B = -1$ otherwise. Now, we will then have the following transformation rules in the chiral formalism.

\begin{itemize}
    \item Under CP.
    \be 
        \ba{c} (\varphi_{\mu})_\alpha \,\xrightarrow{~\text{CP}~}\, 
        -i\eta_B {P_\mu}^{\nu}(\sigma^0)_{\alpha\dot\alpha}(\bar\varphi_{\nu})^{\dot\alpha}~,
        \\[1em]
        (\bar\chi_{\mu})^{\dot\alpha} \,\xrightarrow{~\text{CP}~}\, 
        -i\eta_B {P_\mu}^{\nu}(\sigma^0)^{\dot\alpha\alpha}(\chi_{\nu})_{\alpha}~,
        \ea
    \qquad\text{and}\qquad 
        \ba{c} (\bar\varphi_{\mu})_{\dot\alpha} \,\xrightarrow{~\text{CP}~}\, 
        i\eta_B {P_\mu}^{\nu}(\varphi_{\nu})^{\alpha}(\sigma^0)_{\alpha\dot\alpha}~,    
        \\[1em]
        (\chi_{\mu})^{\alpha} \,\xrightarrow{~\text{CP}~}\,
        i\eta_B {P_\mu}^{\nu}(\bar\chi_{\nu})_{\dot\alpha}(\sigma^0)^{\dot\alpha\alpha}~.
        \ea
        \label{RS.cp}  
    \ee
    \item Under T.
    \be 
        \ba{c} (\varphi_\mu)_\alpha \,\xrightarrow{~\text{T}~}\, 
        -{T_\mu}^\nu(\sigma^0)_{\alpha\beta}(\varphi_\nu)^{\beta}~,
        \\[1em]
        (\bar\chi_\mu)^{\dot\alpha} \,\xrightarrow{~\text{T}~}\, 
        {T_\mu}^\nu(\sigma^0)^{\dot\alpha\dot\beta}(\bar\chi_\nu)_{\dot\beta}~,
    \ea
        \qquad\text{and}\qquad 
        \ba{c} (\bar\varphi_\mu)_{\dot\alpha} \,\xrightarrow{~\text{T}~}\, 
        -{T_\mu}^\nu(\bar\varphi_\nu)^{\dot\beta}(\sigma^0)_{\dot\beta\dot\alpha}~,
        \\[1em]
        (\chi_\mu)^{\alpha} \,\xrightarrow{~\text{T}~}\,
        {T_\mu}^\nu(\chi_\nu)_{\beta}(\sigma^0)^{\beta\alpha}~.
        \ea
        \label{RS.t}
    \ee
    \item Under CPT.
     \be 
        \ba{c} (\varphi_\mu)_\alpha \,\xrightarrow{~\text{CPT}~}\, 
        -i\eta_B {(\sigma^0)_{\alpha}}^{\dot\alpha}(\bar\varphi_\mu)_{\dot\alpha}~,
        \\[1em]
        (\bar\chi_\mu)^{\dot\alpha} \,\xrightarrow{~\text{CPT}~}\, 
        i\eta_B {(\sigma^0)^{\dot\alpha}}_\alpha(\chi_\mu)^{\alpha}~,
        \ea
    \qquad\text{and}\qquad 
        \ba{c} (\bar\varphi_\mu)_{\dot\alpha} \,\xrightarrow{~\text{CPT}~}\, 
        i\eta_B (\varphi_\mu)_{\alpha}{(\sigma^0)^\alpha}_{\dot\alpha}~,
        \\[1em]
        (\chi_\mu)^{\alpha} \,\xrightarrow{~\text{CPT}~}\,
        -i\eta_B (\bar\chi_\mu)^{\dot\alpha}{(\sigma^0)_{\dot\alpha}}^{\alpha}~.
        \ea
    \ee
\end{itemize}
Note that the above transformations are analogous to those presented in Sec.\ref{Chiral.transformations}, with additional factors coming from $B_\mu$. For calculation purposes, it is convenient to analyze tensor currents of the form
    \be \label{general tensor current}
        \chi_\mu\sigma^\delta i\partial_\rho \bar\chi_\nu \quad \text{and} \quad \bar\varphi_\mu \bar\sigma^\delta i\partial_\rho \varphi_\nu \, .
    \ee
And the scalar currents
    \be 
    \begin{aligned} \label{scalar tensor current}
        \eta^{\mu\nu} \bar\varphi_\mu \bar\chi_\nu &\quad \text{and} \quad \eta^{\mu\nu}\chi_\mu \varphi_\nu \, ,\\
        \chi_\mu \sigma^{\mu\nu} \varphi_\nu &\quad \text{and} \quad \bar\varphi_\mu \bar\sigma^{\mu\nu} \bar\chi_\nu \, .
    \end{aligned}
    \ee
The procedure to follow is therefore completely analogous to that outlined in Sec.~\ref{Chiral.transformations}. Of course, additional factors will appear, and in table~\ref{TensorCurr.Sym}, we summarize the results obtained.

\begin{table}[ht]
    \centering
    \caption{Tensor and scalar currents under discrete symmetries}
    \begin{tabular}{|c|c|c|c|}
        \hline
        Current & Under CP & Under Time reversal & Under CPT \\
        \hline
        $\chi_\mu\sigma^\delta i\partial_{\rho}\bar\chi_\nu$ &  $- {P_\mu}^\upsilon{P_\nu}^\tau {P^\delta}_{\omega}{P_\rho}^{\pi}(i\partial_\pi \chi_\tau) \sigma^{\omega}\bar\chi_\upsilon$ & ${T_\mu}^\upsilon {T_\nu}^\tau {T^\delta}_\omega {T_\rho}^\pi \chi_\upsilon \sigma^\omega i\partial_\pi \bar\chi_\tau$ & $-(i\partial_\rho \chi_\nu) \sigma^\delta \bar\chi_\mu$ \\
        $\bar\varphi_\mu\bar\sigma^\delta i\partial_{\rho}\varphi_\nu$ &  $-{P_\mu}^\upsilon{P_\nu}^\tau {P^\delta}_{\omega}{P_\rho}^{\pi} (i\partial_\pi \bar\varphi_\tau)\bar\sigma^{\omega} \varphi_\upsilon$ & ${T_\mu}^\upsilon {T_\nu}^\tau {T^\delta}_\omega {T_\rho}^\pi \bar\varphi_\upsilon \bar\sigma^\omega i\partial_\pi \varphi_\tau$ & $-(i\partial_\rho \bar\varphi_\nu) \bar\sigma^\delta \varphi_\mu$ \\
        $\eta^{\mu\nu} \chi_\mu \varphi_\nu$ & $\eta^{\mu\nu} \bar\varphi_\mu \bar\chi_\nu$ & $\eta^{\mu\nu} \chi_\mu \varphi_\nu$ & $\eta^{\mu\nu} \bar\varphi_\mu \bar\chi_\nu$ \\
        $\eta^{\mu\nu} \bar\varphi_\mu \bar\chi_\nu$ & $\eta^{\mu\nu} \chi_\mu \varphi_\nu$ & $\eta^{\mu\nu} \bar\varphi_\mu \bar\chi_\nu$& $\eta^{\mu\nu} \chi_\mu \varphi_\nu$ \\
        $\chi_\mu \sigma^{\mu\nu} \varphi_\nu$ & $\bar{\varphi}_\mu \bar{\sigma}^{\mu\nu} \bar{\chi}_\nu$ & $\chi_\mu \sigma^{\mu\nu} \varphi_\nu$ & $\bar{\varphi}_\mu \bar{\sigma}^{\mu\nu} \bar{\chi}_\nu$ \\
        $\bar{\varphi}_\mu \bar{\sigma}^{\mu\nu} \bar{\chi}_\nu$ & $\chi_\mu \sigma^{\mu\nu} \varphi_\nu$ & $\bar{\varphi}_\mu \bar{\sigma}^{\mu\nu} \bar{\chi}_\nu$ & $\chi_\mu \sigma^{\mu\nu} \varphi_\nu$ \\
        \hline
    \end{tabular}
    \label{TensorCurr.Sym}
    \end{table}
Using the transformations presented previously, and after a careful treatment of the Lorentz indices as in Sec.~\ref{Chiral.transformations}, we can easily show that most of the terms appearing in the Lagrangian~\eqref{RS Lagrangian.Chiral} (or \eqref{Rarita-Schwigner Lagrangian}) are T, CP, and CPT invariant. In the latter two cases, as well as when analyzing the Dirac Lagrangian, we must consider that total derivatives do not contribute to the Lagrangian. The only term for which it is not so clear what happens is
    \be \label{RS.Free.Massless}
        {\cal L}_{\text{RS}}^{\text{a}} = \frac{1}{2} \epsilon^{\mu\rho\lambda\nu} \bar{\Psi}_\mu \gamma_5 \gamma_\rho \partial_\lambda \Psi_\nu = - \frac{1}{2} \epsilon^{\mu\rho\lambda\nu} (\chi_\mu \sigma_\rho \partial_\lambda \bar\chi_\nu 
        -\bar\varphi_\mu \bar\sigma_\rho\partial_\lambda \varphi_\nu ) \, ,
    \ee
Using the first two results from the table~\ref{TensorCurr.Sym}, and after some algebra, we have that with previous rules for the Minkowski metric and Levi-Civita symbol,
    \be
    \begin{aligned}
        {\cal L}_{\text{RS}}^{\text{a}}  \xrightarrow{~\text{CP}~} & - \frac{i}{2}  \epsilon^{\mu\rho\lambda\nu} {P_\mu}^\upsilon{P_\nu}^\tau {P_\rho}^{\delta}{P_\lambda}^{\pi} \left( - (i\partial_\pi \chi_\tau) \sigma_\delta\bar\chi_\upsilon + (i\partial_\pi \bar\varphi_\tau)\bar\sigma_\delta \varphi_\upsilon \right) \\
        &= + \frac{1}{2} \epsilon^{\mu\rho\lambda\nu} (\chi_\mu \sigma_\rho \partial_\lambda \bar\chi_\nu 
        -\bar\varphi_\mu \bar\sigma_\rho\partial_\lambda \varphi_\nu ) \, ,
    \end{aligned}
    \ee
and 
    \be
    \begin{aligned}
        {\cal L}_{\text{RS}}^{\text{a}}  \xrightarrow{~\text{T}~} & \frac{i}{2} \epsilon^{\mu\rho\lambda\nu} {T_\mu}^\upsilon {T_\nu}^\tau {T_\rho}^\delta {T_\lambda}^\pi \left( \chi_\upsilon \sigma_\delta i\partial_\pi \bar\chi_\tau - \bar\varphi_\upsilon \bar\sigma_\delta i\partial_\pi \varphi_\tau \right) \\
        &= + \frac{1}{2} \epsilon^{\mu\rho\lambda\nu} (\chi_\mu \sigma_\rho \partial_\lambda \bar\chi_\nu 
        -\bar\varphi_\mu \bar\sigma_\rho\partial_\lambda \varphi_\nu ) \, .
    \end{aligned}
    \ee
So this term is CP-odd and T-odd. One might conclude that the general Lagrangian of RS violates CP and T, however, remember that this term comes from the identity: $\gamma^\mu \gamma^\lambda \gamma^\nu = \eta^{\mu\lambda} \gamma^\nu + \eta^{\lambda \nu} \gamma^\mu - \eta^{\mu \nu} \gamma^\lambda - i \epsilon^{\rho \mu \nu \rho} \gamma_\rho \gamma_5$. It is possible to show that the right-hand side of the above identity gives us a term invariant under CP and T, so this identity is valid at the Lagrangian level but not at the level of discrete transformations like CP and T.

As you can see, these terms that violate CP and T cancel each other out when applying CP, and we have concluded that the general RS Lagrangian is CPT invariant as expected since, under proper Lorentz transformations, it transforms as
    \be \label{RS under LT}
        \varphi_\mu (x) \rightarrow \varphi_\mu^\prime (x) = {\Lambda_\mu}^\nu S(\Lambda) \varphi_\nu (\Lambda^{-1} x) \quad \text{and} \quad \bar\chi_\mu (x) \rightarrow \bar\chi_\mu^\prime (x) = {\Lambda_\mu}^\nu [S(\Lambda)^{-1}]^\dagger \bar\chi_\nu (\Lambda^{-1} x)\, .
    \ee
\subsection{On the massless case}
One of the features of local SUSY is the existence of a gauge field associated with the local SUSY, which is called the gravitino. As in the case of ordinary gauge theories, the gravitino couples to a Noether current of SUSY and maintains invariance under the local SUSY transformation. Before SUSY breaking, the gravitino must be a massless particle. The description of massless particles with spin $\frac{3}{2}$ adopted in SUGRA is sometimes called the Rarita-Schwinger Lagrangian~\cite{Van1981supergravity}.
    \be \label{RS.Lagrangian.SUGRA.R}
        {\cal L}_{\text{RS}}^{\textrm{massless}}= -\frac{1}{2} \epsilon^{\mu\rho\lambda\nu}\bar\Psi_{\mu}\gamma_5\gamma_\rho \partial_\lambda \Psi_{\nu} = -\frac{i}{2} \bar\Psi_{\mu} \gamma^{\mu\lambda\nu} \partial_\lambda \Psi_{\nu}~.
    \ee
Here we have used the identity with generalized gamma matrix $i\epsilon^{\mu\rho\lambda\nu} \gamma_5 \gamma_\rho = i\epsilon^{\rho\mu\lambda\nu} \gamma_\kappa \gamma_5 = -\gamma^{\mu\lambda\nu}$. Again this identity is valid at the Lagrangian level since it can be shown that the left-hand side of the identity transformed as a rank 3 tensor is CP and T invariant. Then the free massless RS field satisfies the motion equation
    \be \label{Field equations.RS.SUGRA.R}
       \epsilon ^{\mu \rho \lambda \nu }\gamma _{5}\gamma _\rho \partial_\lambda \Psi_\nu=0 \quad \text{or} \quad  \gamma^{\mu\lambda\nu} \partial_\lambda \Psi_\nu = 0 \, .
    \ee 
One would expect that in the massless limit one would obtain helicity states $\pm \frac{3}{2}$. However, by studying the homogeneous Lorentz group, one can show $SO^{+}(1,3) \rightarrow SO(4) \cong [SU(2) \times SU(2)]/\mathbb{Z}_2$, so at the level of Lie algebra $\mathfrak{so}(1,3) \cong \mathfrak{su}(2) \otimes \mathfrak{su}(2)$. Therefore, we can describe a field as an irrep $(j_L, j_R)$. A field of this type can be formed only from the annihilation operators for a massless particle of helicity $\sigma^{+}$ and the creation operators for the antiparticle of helicity $\sigma^{-}$, where
    \be
        \sigma^{\pm} = \pm (j_R - j_L) \, .
    \ee
Then for the representation, $\left( 1, \frac{1}{2} \right) \oplus \left( \frac{1}{2},1 \right)$ the helicity states in massless limit are
    \be
        h^{\pm} = \pm \frac{1}{2} \, .
    \ee
Considering that the massless limit is equivalent to the high energy limit, these same helicity states should be obtained, and in fact, they correspond to goldstino states found in~\cite{Diaz2017helicity}. This then suggests that the description of a particle with helicity $\pm \frac{3}{2}$ transforms under the complete Lorentz group as the representation $\left(\frac{3}{2},0 \right) \oplus \left( 0, \frac{3}{2} \right)$ (plus constraints). It can be easily verified that this representation comes from $[(1,0) \oplus (0,1)] \otimes \left(\frac{1}{2},0 \right) \oplus \left[\left( 0, \frac{1}{2} \right) \right]$. This prescription has been worked with in the literature, with the general description of fields described by Weinberg-Joos states $(j,0) \oplus (0,j)$, where the Lorentz invariants are independent of momentum and do not provide wave equations and the genuine wave equations are obtained by conditioning the Lorentz tensors to satisfy the Klein-Gordon equation. In doing so, one always ends up with wave equations and associated Lagrangians that are second order in momenta. For instance, a spin-$3/2$ particle that transforms as $\left(\frac{3}{2},0 \right) \oplus \left( 0, \frac{3}{2} \right)$ is conveniently described by a second order Lagrangian in the basis of the fully antisymmetric second rank Lorentz tensor-spinor $\Psi_{[\mu\nu]}$ \cite{Delgado2015bosonic}. But clearly the Lagrangian in that case does not correspond to the RS Lagrangian, the alternative used in SUGRA it is obtained by noting that $\gamma_\mu \gamma^{\mu\lambda\nu} \partial_\lambda \Psi_\nu = 2\gamma^{\rho\nu} \partial_\lambda \Psi_\nu = 0$, which implies that 
    \be \label{FE.Alt}
        \gamma^{\nu \lambda} \partial_{\nu} \Psi_\lambda =0 \, ,
    \ee
This is an alternative form of the equation of motion, equivalent to \eqref{Field equations.RS.SUGRA.R}, but which cannot be obtained directly from an action. Now, we can rewrite
    \be
        ( - \gamma^\lambda\gamma^\nu +  \eta^{\nu\lambda}) \partial_\lambda \Psi_\nu = \partial \cdot \Psi - \slashed{\partial} \gamma \cdot \Psi = 0 \, .
        \label{E.1}
    \ee
or, notice that we choose $\gamma^{\nu\lambda}\partial_\lambda \Psi_\nu = 0$, and $\gamma^{\lambda\nu}\partial_\nu \Psi_\lambda =0$, taking the sum the field equation becomes to
    \be \label{E.2}
        \gamma^\nu \gamma^\lambda ( \partial_\lambda \Psi_\nu - \partial_\nu \Psi_\lambda) \Rightarrow \slashed{\partial} \Psi_\nu - \partial_\nu \gamma \cdot \Psi = 0 \, .
    \ee
The equations \eqref{E.1} and \eqref{E.2} are those that appear in the literature \cite{Freedman2012supergravity}. And in the $\gamma_\mu \Psi^\mu = 0$ gauge, as
    \be \label{Constrains.RS original.2}
        \slashed{\partial} \Psi_\mu = 0 , \quad \partial_\mu \Psi^\mu = 0 , \quad \gamma^\mu \Psi_\mu \big|_{\text{gf}} = 0 \, .. 
    \ee
The third equation corresponds to the gauge choice that fixes the freedom $\delta_\eta \Psi_\mu = \partial_\mu \eta$, where the symbol $\big|_{\text{gf}}$ reflects this. In the massive RS system, $\partial^\mu \Psi_\mu = 0$ is a consistency condition of the field equations, hence \eqref{Constrains.RS original.2} can be obtained from the massless limit of the original RS equations, however, \eqref{Constrains.RS original.2} can not be obtained by direct variation of the massless action. 

Again, as in the case of the gravitino in order to obtain the degrees of freedom for a massless RS field, it is better to use a gauge analogous to the Coulomb gauge. For this purpose, let's apply $\partial_\lambda$ to \eqref{FE.Alt} and antisymmetrize in $\lambda \nu$ to obtain $\slashed{\partial}\left(\partial_{\lambda} \Psi_{\nu}-\partial_{\nu} \Psi_{\lambda}\right) = 0$. We need to fix the gauge, so we impose the non-covariant condition
    \be
        \gamma^j \Psi_j = 0 \, ,
    \ee
We use the equivalent form \eqref{FE.Alt} of the field equations. The $\nu=0$ and $\nu \rightarrow i$ components are
    \be
    \begin{aligned}
        \gamma^j \partial_j \Psi_{0}-\partial_j \gamma^j \Psi_j &=0 \, ,\quad 
        \gamma \cdot \partial \Psi_j -\partial_j \gamma \cdot \Psi &=0 \, .
    \end{aligned}
    \ee
Using the gauge condition one can see that $\nabla^{2} \Psi_{0}=0$, so $\Psi_{0}=0$. The spatial-like components $\Psi_i$ then satisfy the Dirac-like equation
    \be
        \gamma \cdot \partial \Psi_j =0 \, ,
        \label{Dirac-like.spatial}
    \ee
which is a time evolution equation. However, there is an additional constraint, $\partial^{i} \Psi_{i}=0$, obtained by contracting \eqref{Dirac-like.spatial} with $\gamma^{i}$. Thus from the gauge condition and the equation of motion, we find $12$ independent constraints on the initial data, namely in chiral fermion formalism
    \be \label{Chiral.RS.Massless}
    \begin{aligned}
        \ba{rclcrcl}
        (\sigma^j )_{\alpha \dot\alpha} (\bar\chi_j)^{\dot\alpha} &=& 0 &&
        (\bar\sigma^j )^{\dot\alpha \alpha} (\varphi_j)_{\alpha} &=& 0  \nonumber\\
        (\bar\chi_0)^{\dot\alpha} &=& 0 &&
        (\varphi_0)_{\alpha} &=& 0 \nonumber\\
        \partial_j (\varphi^j)_\alpha &=& 0 && \partial_j(\bar\chi^j)^{\dot\alpha} &=& 0 \, .
        \ea
    \end{aligned} \tag{6.60}
    \ee
These constraints imply that there are only $4$ initial components of $\Psi_j$ to be specified. The time derivatives are already determined by the Dirac-like equation \eqref{Dirac-like.spatial}. Hence there are $4$ classical degrees of freedom for the RS gauge field in $4$-dimensional Minkowski space. The on-shell degrees of freedom are half this number. In dimension $d=4$, with Majorana conditions, we find the two states expected for a massless particle. A treatment for the quantization of the RS field has been proposed by considering the helicity $\pm \frac{3}{2}$ states as representations of $SO(2) $~\cite{Freedman2012supergravity}, this comes from studying the so-called {\it Little group} for massless particles, $ISO(2) \cong SO(2) \ltimes \mathbb{R}^2$. Here $SO(2)$ describes the physical helicity states of the particle, but this does not provide an adequate description of the transformations under the Lorentz group that we establish here. To avoid this kind of complications as to whether there is a direct correspondence or not, there is a formalism in which we only care about the physical states of helicity and we can determine amplitudes in processes involving massless gravitinos, an example can be seen in~\cite{ChakrabortyJonathanLorenzo2024}.

The term \eqref{RS.Free.Massless} can be present in the Lagrangian describing spin-$\frac{3}{2}$ (massless or massless). In such a case both the CP and time reversal symmetry, T, are odd, thus breaking these symmetries. However, one can choose formulations where these symmetries are always respected.
\section{CPT AND LORENTZ INVARIANCE VIOLATION: MODEL EXTENSION} \label{Model Extension}
Effective field theory (EFT) is useful for describing observable signals of Lorentz violation. Any realistic EFT must contain the Lagrangian densities for both general relativity and the standard model. By also adding all terms involving operators for Lorentz violation that are scalar under coordinate transformations, one obtains an effective field theory called the standard model extension (SME).

Non-zero Lorentz violation SME coefficients can arise from two main mechanisms: spontaneous and explicit Lorentz violation. In the case of spontaneous violation, the coefficients result from underlying dynamics and act as fields that influence the system through action variation. Conversely, explicit Lorentz violation implies the coefficients are fixed functions of spacetime with no dynamical impact. Importantly, the geometry of Riemann-Cartan spaces, including standard Riemannian limits, is incompatible with explicit Lorentz violation \cite{Kostelecky2004gravity}. 

The description of the gravitational field in linearized gravity that is useful to describe the graviton has been presented in the section~\ref{W.gravitational field}, we could notice that its description is different from the description of spin-$\frac{3}{2}$ fermions that is done through a Dirac-like equation, and the formulation described in the section~\ref{Rs. theories}. We are interested in exploring the extension of these free field theories (without interactions) to include terms that may violate CPT symmetry and therefore Lorentz symmetry. Spontaneous violation of Lorentz symmetry is a topic that has been studied quite a bit in recent years, particularly in the fermionic sector for spin-$\frac{1}{2}$ particles~\cite{Kostelecky1991cpt,Kostelecky1995cpt}. Thus, we seek to extrapolate the SME to the inclusion of spin-$\frac{3}{2}$ particles,
    \be
        {\cal L}_{\text{EH}}^{\text{Ex}} = {\cal L}_{\text{EH}}^{\text{L}} - {{\cal L}_{\text{LV}}^{\text{L}}} \quad \text{and} \quad {\cal L}_{\text{RS}}^{\text{Ex}} = {\cal L}_{\text{RS}} - {\cal L}_{\text{LV}}^\prime
    \ee
We are interested in studying possible ways that ${\cal L}_{\text{LV}}^{\text{L}}$ and ${\cal L}_{\text{LV}}^\prime$ could arise as effective contributions of spontaneous CPT violation in a complete theory. Thinking that with this we are going towards completing the framework in which there is a LIV in SUGRA.
\subsection{CPT and Lorentz invariance violation in linearized gravity} \label{LIV.LG}
The pure gravity limit assumes Riemannian geometry, where the usual EH action is complemented by all possible coordinate-independent terms that involve the metric $g_{\mu\nu}$. This construction produces an effective field theory that captures all realistic violations of Lorentz invariance in gravity~\cite{Kostelecky2004gravity, Bluhm2005spontaneous}. However, the challenges of working with a nonlinear theory like general relativity are further compounded by the inclusion of Lorentz violations, complicating systematic studies. Many of these complications can be avoided by working within the LG framework. 
Lorentz-violating term in the LG action is formed from the contraction of a conventional tensor operator with a Lorentz-violating tensor coefficient. The Lorentz-violating piece ${\cal L}^{\text{L}}_{\text{LV}}$ is constructed by combining Lorentz-violating coefficients with gravitational field operators to produce individual terms that are observer-invariant for both local Lorentz transformations and general coordinate transformations. Terms are classified according to the mass dimension $d$ of the operator in natural units. We can think that the higher dimension terms $d$ represent higher order corrections to conventional physics. Taking into account the compact way of writing the LG action presented in \eqref{EH.L.Lichnerowicz} or \eqref{Graviton Lagrangian}, we can see that a generic term that violates Lorentz symmetry takes the form
    \be
        {\cal L}_{\text{LV}}^{\text{L}} = \frac{1}{4} h_{\mu\nu} \hat{K}^{(d)\mu\nu \rho \omega} h_{\rho\omega} \, .
    \ee
Varying the action, it is straightforward to see that a non-zero contribution must satisfy  $(\hat{K}^{(d)(\mu\nu) (\rho \omega)} + \hat{K}^{(d)(\rho\omega)(\mu\nu)}) \neq 0$. Considering the usual gauge transformation valid in general relativity, $\delta h_{\mu\nu} = \partial_\mu \xi_\nu + \partial_\nu \xi_\mu$ , and which allows us to adequately describe gravitational fields because it helps us to eliminate nonphysical components, $(\hat{K}^{(d)(\mu\nu) (\rho \omega)} + \hat{K}^{(d)(\rho\omega)(\mu\nu)}) \partial_\nu = 0$. There are three general classes of Lorentz-violating operators in linearized gravity~\cite{Kostelecky2018lorentz}. According to the transformations introduced in subsection \ref{LG.Transformations}, we have that under discrete symmetries
 \begin{table}[ht] \label{Lorentz.V.CPT}
    \centering
    \caption{Operators spontaneously violating Lorentz invariance with gauge conservation~\cite{Kostelecky2018lorentz}}
    \begin{tabular}{|c|c|c|c|c|}
        \hline
        Operator & Young Tableaux & Under CPT transformation & $d$ & No. of independent components \\
        \hline
        $s^{(d) \mu\rho \circ \nu \omega \circ \circ^{d-4}}$ & $\ytableaushort{\mu \nu \cdots , \rho \omega, \circ \circ }$ & even & even, $d\geq 4$ & $(d-3)(d-2)(d+1)$ \\ \hline 
        $q^{(d)\mu\rho \circ \nu \circ \omega \circ \circ^{d-5}}$ & $\ytableaushort{\mu \nu \omega \cdots , \rho \circ \circ, \circ }$ & odd & odd, $d\geq 3$ & $\frac{5}{2}(d-4)(d-1)(d+1)$ \\
        \hline
        $k^{(d)\mu \circ \rho \circ \nu \circ \omega \circ \circ^{d-6}}$ & $\ytableaushort{\mu \nu \rho \omega \cdots , \circ \circ \circ \circ}$ & even & even, $d\geq 6$ & $\frac{5}{2}(d-5)(d+1)$ \\
        \hline
    \end{tabular}
    \end{table}

\begin{table}[ht] \label{Lorentz.V.CP.T}
    \centering
    \caption{Operators spontaneously violating Lorentz invariance under discrete symmetries}
    \begin{tabular}{|c|c|c|c|c|}
        \hline
        Lagrangian term & Under CP transformation & Under Time reversal \\
        \hline
        $h_{\mu\nu} s^{(d)\mu\rho \circ \nu \omega \circ \circ^{d-4}} h_{\rho\omega}$ & $\prod_{j=1}^{d-4} {P_{\varepsilon_j}}^{\upsilon_j} h_{\mu\nu }S^{\mu\rho \circ \nu \omega \circ \circ^{d-4}_{\upsilon_j}} h_{\rho\omega}$ &  $\prod_{j=1}^{d-4} {T_{\varepsilon_j}}^{\upsilon_j} h_{\mu\nu}S^{\mu\rho \circ \nu \omega \circ \circ^{d-4}_{\upsilon_j}}h_{\rho\omega}$ \\ \hline
        $h_{\mu\nu} q^{(d)\mu\rho \circ \nu \circ \omega \circ \circ^{d-5}} h_{\rho\omega}$ & $\prod_{j=1}^{d-5} {P_{\varepsilon_j}}^{\upsilon_j} h_{\mu\nu } q^{(d)\mu\rho \circ \nu \circ \omega \circ \circ^{d-5}} h_{\rho\omega}$  & $\prod_{j=1}^{d-5} {T_{\varepsilon_j}}^{\upsilon_j} h_{\mu\nu} q^{(d)\mu\rho \circ \nu \circ \omega \circ \circ^{d-5}} h_{\rho\omega}$ \\ \hline
        $h_{\mu\nu} k^{(d)\mu \circ \rho \circ \nu \circ \omega \circ \circ^{d-6}} h_{\rho\omega}$ & $\prod_{j=1}^{d-6} {P_{\varepsilon_j}}^{\upsilon_j} h_{\mu\nu } k^{(d)\mu \circ \rho \circ \nu \circ \omega \circ \circ^{d-6}} h_{\rho\omega}$ & $\prod_{j=1}^{d-6} {T_{\varepsilon_j}}^{\upsilon_j} h_{\mu\nu} k^{(d)\mu \circ \rho \circ \nu \circ \omega \circ \circ^{d-6}} h_{\rho\omega}$  \\
        \hline
    \end{tabular}
    \end{table}
Here we have used the short-notation ${}^{\circ^{d-m}_{\upsilon_j}} \equiv \partial_{\upsilon_1} \partial_{\upsilon_2} \cdots \partial_{\upsilon_{d-m}}$ with $m=4,5,6$. An inspection of the transformations tells us that the terms violate CP and T in some of their components, and in cases where we have an even number of derivatives the relation preserves CPT by relation ${P_{\varepsilon_j}}^{\upsilon_j} {T_{\upsilon_j}}^{\epsilon_j} = - {\delta_{\varepsilon_j}}^{\epsilon_j}$ for $j \in \{ 1, 2, \dots , d-m\}$, otherwise, we obtain a CPT-odd operator, of course this is in agreement with the fact that there are no factors affecting the field $h_{\mu\nu}$ under CPT symmetry. The particle sectors of the SME have received significant attention over the past two decades~\cite{Kostelecky2011data}, but fewer studies have focused on Lorentz violations in gravity. For example, one tests of Lorentz invariance in the gravitational sector of the SME involve studies of gravitational `Cerenkov radiation~\cite{Kostelecky2015constraints}, where a specific case of $s^{(d)\mu\rho \circ \nu \omega \circ \circ^{d-4}}$ (for $d=4$) is examined that is which is equivalent to a simple replacement in the Lichnerowicz operator $\eta_{\mu\nu} \rightarrow s_{\mu\nu}$ (with $s_{\mu\nu} = s_{\nu\mu}$).
\subsection{CPT and Lorentz invariance violation in Rarita-Schwinger theory} \label{LIV.RS}
To date, string theory constitutes the only class of (gauge) theories in four or more dimensions that are quantum-consistent, dynamically invariant under Poincaré symmetry, and known to admit an explicit mechanism \cite{Kostelecky1991cpt,Kostelecky1995cpt} for spontaneous Lorentz invariance violation (LIV). In particle field theory, {\it spontaneous symmetry breaking} happens when the symmetries of the Lagrangian are not preserved by the ground state of the theory. This occurs if the naive perturbative vacuum—where all fields have zero expectation value is unstable. In the true vacuum, some fields acquire nonzero expectation values, breaking any symmetry of the Lagrangian that does not leave these values invariant.

A similar idea applies in {\it covariant string field theory}. When Lorentz symmetry is spontaneously broken, the relevant fields are tensor fields that transform nontrivially under the homogeneous Lorentz group. We denote these generically as $T_{I}(x)$, where $I = \mu\nu \dots \rho$ represents one or more Lorentz vector indices. If any quadratic term in $T_{I}$ has a negative coefficient in the potential, some components of $T$ develop nonzero expectation values, causing the Lorentz symmetry to break spontaneously.

For decades, researchers have explored the possibility of {\it CPT symmetry breaking} through the spontaneous breaking of Lorentz symmetry in the spin-$\frac{1}{2}$ fermion sector. This is done by introducing effective operators from string field theory. In fact, any term that violates CPT symmetry in an effective four-dimensional theory must have mass dimension four~\cite{Colladay1997cpt}. The operators arise from an effective theory having the general form
    \be
        {\cal L}^{\prime} \supset \frac{\lambda}{M^{k}}\langle T\rangle \cdot \bar{\Psi} \Gamma(i \partial)^{k} \Psi +\text { h.c.} \, ,
        \label{P.Ref}
    \ee
the parameter $\lambda$ is a dimensionless coupling constant, $(i\partial)^k$ represents $k$ four-derivatives acting in some combination on the fermion fields, and $\Gamma$ represents
some gamma-matrix structure, where the expectations of the tensors $T_{I}$, $\la T \ra$, are assumed to violate Lorentz laws and possibly the CPT symmetry, here, we have omitted the Lorentz indices, and $M$ is large mass scale, which is assumed to be the Planck mass~\cite{Bluhm2005spontaneous}. Terms with $\leq 3$ and with more quadratic fermion factors also appear, but these are further suppressed. To introduce fermions into string theories, supersymmetry must be included, and therefore, supersymmetric partners of gravitons, the gravitinos, are needed. Since the free RS theory described by the Lagrangian is renormalizable by making a convenient choice of the propagator, the terms to explore for these fermions that would not affect the renormalization are those for which $k \leq 1$, like spin-$\frac{1}{2}$ fermions. A general term that violates Lorentz invariance and/or CPT symmetry in RS theory has the following general form
    \be
        \bar \Psi_\mu \tilde{\Lambda}^{\mu\nu}_{\text{k}} \Psi_\nu \quad \text{with} \quad [\tilde{\Lambda}^{\mu\nu}_{\text{k}}] = {\rm mass} \, .
    \ee
Taking into account that in general the RS Lagrangian must be Hermitian by imposing this condition we find that the terms that violate Lorentz and/or CPT must fulfill the condition $\tilde{\Lambda}^{\mu\nu}_k = \gamma^0 (\tilde{\Lambda}^{\nu\mu}_k)^\dagger \gamma^{0}$~\footnote{The same condition for constructing the most general RS Lagrangian}. 

Considering that the gamma matrices and combinations of them satisfy: $\gamma^\lambda = \gamma^0 (\gamma^\lambda)^\dagger \gamma^0$, $\gamma^\lambda = \gamma^0 (\gamma_5\gamma^\lambda)^\dagger \gamma^0$, $i\gamma_5 = \gamma^0 (i\gamma_5)^\dagger \gamma^0$, 
It is straightforward to notice through an inspection of \eqref{RS.Op} and \eqref{Rarita-Schwigner Lagrangian} that the possible low-order operators $\tilde{\Lambda}^{\mu\nu}_{\text{k}}$ must satisfy the hermiticity conditions. All coefficients of order $k=0$ can be real and symmetric in the indices $\mu,\nu$. If they are antisymmetric, they must be accompanied by an ``$i$'' to conform to the hermiticity condition.
    \be
        \tilde{\Lambda}^{\mu\nu}_0 = \begin{cases}
            {{C_s}^{(\mu\nu)}}_\lambda \gamma^\lambda  \\
            i{{C_a}^{[\mu\nu]}}_\lambda \gamma^\lambda  \\
            {D_s}^{(\mu\nu)\lambda} \gamma_5\gamma_\lambda\\
            i{D_a}^{[\mu\nu]\lambda} \gamma_5\gamma_\lambda\\
            i E^{(\mu\nu)} \gamma_5 \\
            G^{(\mu\nu)} \\
        \end{cases} \quad \text{and} \quad 
        \tilde{\Lambda}^{\mu\nu}_1 = \Omega^{\mu\nu \lambda} i\partial_\lambda = \begin{cases}
            q^{(\mu\nu)}_1 \gamma^\lambda i\partial_\lambda \\
            q^{(\mu\nu)}_2 \gamma_5 \gamma^\lambda i\partial_\lambda \\
            \eta^{\mu\nu} f^\lambda_1 i\partial_\lambda \\
            -if_2^\lambda \Sigma^{\mu\nu} (i\partial_\lambda) \\
            \eta^{\mu\nu} g^\lambda_1 \gamma_5 i\partial_\lambda \\ 
            -ig^\lambda_2 \gamma_5 \Sigma^{\mu\nu} (i \partial_\lambda )
        \end{cases} \, .
    \ee
Here we have used the conventional notation that $( , )$ indicates that the indices are symmetrized and that $[,]$ indicates that they are antisymmetrized. The coefficients are real at order $0$ and have mass dimensions, while at order $1$ are dimensionless, notice that $f_j^\lambda , g_j^\lambda$ are axial vectors for $j=1,2$. One might think that we are missing more terms at order $k=1$, however, note that we have taken into account the identities $\gamma^{\mu\nu} = \eta^{\mu\nu} - i \Sigma^{\mu\nu}$ and $\gamma^\mu \gamma^\lambda \gamma^\nu = \eta^{\mu\lambda} \gamma^\nu + \eta^{\lambda\nu} \gamma^\mu - \eta^{\mu\nu} \gamma^\lambda - i\epsilon^{\rho\mu\lambda\nu} \gamma_\rho \gamma_5$ which are just the ones we used to establish \eqref{Rarita-Schwigner Lagrangian} from \eqref{RS.Op}. Now using the chiral formalism, we can write the Lorentz invariance violating (LIV) terms according to the transformations \eqref{RS under LT} in chiral formalism
    \be
        \bar\Psi_\mu \tilde{\Lambda}^{\mu\nu}_0 \Psi_\nu = \begin{cases}
            {{C_s}^{(\mu\nu)}}_\lambda \bar\Psi_\mu \gamma^\lambda \Psi_\nu ={{C_s}^{(\mu\nu)}}_\lambda (\chi_\mu \sigma^\lambda \bar{\chi}_\nu + \bar{\varphi_\mu} \bar\sigma^\lambda \varphi_\nu) \\
            i{{C_a}^{[\mu\nu]}}_\lambda \bar\Psi_\mu \gamma^\lambda \Psi_\nu = i{{C_a}^{[\mu\nu]}}_\lambda (\chi_\mu \sigma^\lambda \bar{\chi}_\nu + \bar{\varphi_\mu} \bar\sigma^\lambda \varphi_\nu) \\
            {D_s}^{(\mu\nu)\lambda} \bar\Psi_\mu \gamma_5 \gamma_\lambda \Psi_\nu = {D_s}^{(\mu\nu)\lambda} (\bar{\varphi_\mu} \bar\sigma_\lambda \varphi_\nu - \chi_\mu \sigma_\lambda \bar{\chi}_\nu) \\
            iE^{\mu\nu} \bar\Psi_\mu \gamma_5 \Psi_\nu = iE^{\mu\nu} (\bar\varphi_\mu \bar\chi_\nu - \chi_\mu \varphi_\nu) \\
            G^{\mu\nu} \bar\Psi_\mu \Psi_\nu = G^{\mu\nu} (\bar\varphi_\mu \bar\chi_\nu + \chi_\mu \varphi_\nu) 
        \end{cases} \, ,
    \ee
and 
    \be
        \bar\Psi_\mu \tilde{\Lambda}^{\mu\nu}_1 \Psi_\nu = \begin{cases}
            q^{\mu\nu}_1 \bar\Psi_\mu \gamma^\lambda i\partial_\lambda \Psi_\nu = q^{\mu\nu}_1 (\chi_\mu \sigma^\lambda i\partial_\lambda \bar{\chi}_\nu + \bar{\varphi_\mu} \bar\sigma^\lambda i\partial_\lambda \varphi_\nu ) \\
            q^{\mu\nu}_2 \bar\Psi_\mu \gamma_5 \gamma^\lambda i\partial_\lambda \Psi_\nu = q^{\mu\nu}_2 (\bar\varphi_\mu \bar\sigma^\lambda i\partial_\lambda \varphi_\nu - \chi_\mu \sigma^\lambda i\partial_\lambda \bar\chi_\nu) \\
            \eta^{\mu\nu} f^\lambda_1 \bar\Psi_\mu i\partial_\lambda \Psi_\nu = \eta^{\mu\nu} f^\lambda_2 (\chi_\mu i\partial_\lambda \varphi_\nu + \bar\varphi_\mu i\partial_\lambda \bar\chi_\nu) \\
            -if_2^\lambda \bar\Psi_\mu \Sigma^{\mu\nu} i\partial_\lambda \Psi_\nu =  2f^\lambda_2 (\chi_\mu \sigma^{\mu\nu} i\partial_\lambda \varphi_\nu + \bar\varphi_\mu \bar\sigma^{\mu\nu} i\partial_\lambda \bar\chi_\nu) \\
            \eta^{\mu\nu} g^\lambda_1 \bar\Psi_\mu \gamma_5 i\partial_\lambda \Psi_\nu = \eta^{\mu\nu} g^\lambda_1 (\bar\varphi_\mu i\partial_\lambda \bar\chi_\nu - \chi_\mu i\partial_\lambda \varphi_\nu )\\ 
            -ig^\lambda_2 \bar\Psi_\mu \gamma_5 \Sigma^{\mu\nu} i \partial_\lambda \Psi_\nu = 2g^\lambda_2 (\bar\varphi_\mu \bar\sigma^{\mu\nu} i\partial_\lambda \bar\chi_\nu - \chi_\mu \sigma^{\mu\nu} i\partial_\lambda \varphi_\nu)
        \end{cases} \, .
    \ee
In order to establish a classification of these operators under discrete symmetries, it is convenient to establish the following transformations for the tensor currents (see table \ref{TensorCurr.LV})
    \begin{table}[ht]
    \centering
    \caption{Tensor currents under discrete symmetries}
    \begin{tabular}{|c|c|c|c|}
        \hline
        Current & Under CP & Under Time reversal & Under CPT \\
        \hline
        $\chi_\mu\sigma^\delta \bar\chi_\nu$ &  $- {P_\mu}^\upsilon{P_\nu}^\tau {P^\delta}_{\omega} \chi_\tau \sigma^{\omega}\bar\chi_\upsilon$ & $-{T_\mu}^\upsilon {T_\nu}^\tau {T^\delta}_\omega \chi_\upsilon \sigma^\omega \bar\chi_\tau$ & $-\chi_\nu \sigma^\delta \bar\chi_\mu$ \\
        $\bar\varphi_\mu\bar\sigma^\delta \varphi_\nu$ &  $-{P_\mu}^\upsilon{P_\nu}^\tau {P^\delta}_{\omega} \bar\varphi_\tau \bar\sigma^{\omega} \varphi_\upsilon$ & $-{T_\mu}^\upsilon {T_\nu}^\tau {T^\delta}_\omega \bar\varphi_\upsilon \bar\sigma^\omega \varphi_\tau$ & $-  \bar\varphi_\nu \bar\sigma^\delta \varphi_\mu$ \\
        $\chi_\mu \varphi_\nu$ & ${P_\mu}^{\upsilon} {P_\nu}^{\tau}  \chi_\tau \varphi_\upsilon$ & $ {T_\mu}^\upsilon {T_\nu}^\tau \chi_\upsilon \varphi_\tau$ & $\bar{\varphi}_\nu \bar{\chi}_\mu$ \\
        $\bar\varphi_\mu \bar\chi_\nu$ & ${P_\mu}^{\upsilon} {P_\nu}^{\tau} \bar\varphi_\tau \bar\chi_\upsilon$ & $ {T_\mu}^\upsilon {T_\nu}^\tau \bar\varphi_\upsilon \bar\chi_\tau $ & $\chi_\nu \varphi_\mu$ \\
        $\chi_\mu i\partial_\lambda \varphi_\nu$ & ${P_\mu}^{\upsilon} {P_\nu}^{\tau} {P_\lambda}^\omega (i\partial_\omega \chi_\tau) \varphi_\upsilon$ & $- {T_\mu}^\upsilon {T_\nu}^\tau {T_\lambda}^\omega \bar\varphi_\upsilon i\partial_\omega \bar\chi_\tau $ & $(i\partial_\lambda \bar{\varphi}_\nu) \bar{\chi}_\mu$ \\
        $\bar\varphi_\mu i\partial_\lambda \bar\chi_\nu$ & ${P_\mu}^{\upsilon} {P_\nu}^{\tau} {P_\lambda}^\omega (i\partial_\omega \bar\varphi_\tau) \bar\chi_\upsilon$ & $- {T_\mu}^\upsilon {T_\nu}^\tau {T_\lambda}^\omega \chi_\upsilon i\partial_\omega \varphi_\tau$ & $(i\partial_\lambda \chi_\nu )\varphi_\mu$ \\
        $\chi_\mu \sigma^{\mu\nu} i\partial_\lambda \varphi_\nu$ & ${P_\lambda}^\omega ( i\partial_\omega \bar{\varphi}_\mu ) \bar{\sigma}^{\mu\nu} \bar{\chi}_\nu$ & $-{T_\lambda}^\omega \chi_\mu \sigma^{\mu\nu} i\partial_\omega \varphi_\nu$ & $-(i\partial_\lambda \bar\varphi_\mu) \bar\sigma^{\mu\nu} \bar\chi_\nu$ \\
        $\bar{\varphi}_\mu \bar{\sigma}^{\mu\nu} i\partial_\lambda \bar{\chi}_\nu$ & ${P_\lambda}^\omega (i\partial_\omega \chi_\mu) \sigma^{\mu\nu} \varphi_\nu$ & $-{T_\lambda}^\omega \bar{\bar\varphi}_\mu \bar\sigma^{\mu\nu} i\partial_\omega \bar{\bar\chi}_\nu$ & $ - (i\partial_\lambda \chi_\mu) \sigma^{\mu\nu} \varphi_\nu$ \\
        \hline
    \end{tabular}
    \label{TensorCurr.LV}
    \end{table}
    
Using these results and considering that the total derivatives do not contribute to the action we have the following transformation rules in the Table.~\ref{Lorentz-V.CPT}.

    \begin{table}[ht] 
    \centering
    \caption{LIV terms in RS theory}
    \begin{tabular}{|c|c|c|c|c|}
        \hline
        Order & Operator & CP & T & CPT   \\ \hline
        \multirow{6}{*}{$k=0$}
        & ${{C_s}^{(\mu\nu)}}_\lambda \bar\Psi_\mu \gamma^\lambda \Psi_\nu $ & $- {{C_s}^{(\mu\nu)}}_\lambda {P_\mu}^\upsilon{P_\nu}^\tau {P^\lambda}_{\omega} \bar\Psi_\tau \gamma^\omega \Psi_\upsilon$ & $- {{C_s}^{(\mu\nu)}}_\lambda {T_\mu}^\upsilon{T_\nu}^\tau {T^\lambda}_{\omega} \bar\Psi_\upsilon \gamma^\omega \Psi_\tau$ & odd  \\ 
        \cline{2-5} & $i{{C_a}^{[\mu\nu]}}_\lambda \bar\Psi_\mu \gamma^\lambda \Psi_\nu $ & $- i{{C_a}^{[\mu\nu]}}_\lambda {P_\mu}^\upsilon{P_\nu}^\tau {P^\lambda}_{\omega} \bar\Psi_\tau \gamma^\omega \Psi_\upsilon$ & $i {{C_a}^{[\mu\nu]}}_\lambda {T_\mu}^\upsilon{T_\nu}^\tau {T^\lambda}_{\omega} \bar\Psi_\upsilon \gamma^\omega \Psi_\tau$ & odd  \\ 
        \cline{2-5} & ${D_s}^{(\mu\nu)\lambda} \bar\Psi_\mu \gamma_5 \gamma_\lambda \Psi_\nu$ & $ - {D_s}^{(\mu\nu)\lambda} {P_\mu}^\upsilon{P_\nu}^\tau {P_\lambda}^{\omega} \bar\Psi_\tau \gamma_5 \gamma_\omega \Psi_\upsilon$ & $- {D_s}^{(\mu\nu)\lambda} {T_\mu}^\upsilon{T_\nu}^\tau {T_\lambda}^{\omega} \bar\Psi_\upsilon \gamma_5 \gamma_\omega \Psi_\tau$ & odd \\
        \cline{2-5} & $i{D_a}^{[\mu\nu]\lambda} \bar\Psi_\mu \gamma_5 \gamma_\lambda \Psi_\nu$ & $ - i{D_a}^{[\mu\nu]\lambda} {P_\mu}^\upsilon{P_\nu}^\tau {P_\lambda}^{\omega} \bar\Psi_\tau \gamma_5 \gamma_\omega \Psi_\upsilon$ & $i{D_a}^{[\mu\nu]\lambda} {T_\mu}^\upsilon{T_\nu}^\tau {T_\lambda}^{\omega} \bar\Psi_\upsilon \gamma_5 \gamma_\omega \Psi_\tau$ & odd \\
        \cline{2-5} & $i E^{(\mu\nu)} \bar\Psi_\mu \gamma_5 \Psi_\nu $ & $ -i E^{(\mu\nu)} {P_\mu}^\upsilon {P_\nu}^\tau \bar\Psi_\tau \gamma_5 \Psi_\upsilon$ & $ -i E^{(\mu\nu)} {T_\mu}^\upsilon {T_\nu}^\tau \bar\Psi_\upsilon \gamma_5 \Psi_\tau$ & even \\ 
        \cline{2-5} & $G^{(\mu\nu)} \bar\Psi_\mu \Psi_\nu $ & $ G^{(\mu\nu)} {P_\mu}^\upsilon {P_\nu}^\tau \bar\Psi_\tau \Psi_\upsilon$ & $G^{(\mu\nu)} {T_\mu}^\upsilon {T_\nu}^\tau \bar\Psi_\upsilon \Psi_\tau$ & even \\
        \hline
        \multirow{6}{*}{$k=1$} & $q^{(\mu\nu)}_1 \bar\Psi_\mu \gamma^\lambda i\partial_\lambda \Psi_\nu \quad$ &  $q^{(\mu\nu)}_1 {P_\mu}^\upsilon{P_\nu}^\tau \bar\Psi_\upsilon \gamma^\lambda i\partial_\lambda \Psi_\tau $ & $q^{(\mu\nu)}_1 {T_\mu}^\upsilon{T_\nu}^\tau \bar\Psi_\upsilon \gamma^\lambda i\partial_\lambda \Psi_\tau$ & even \\
        \cline{2-5} & $ q^{(\mu\nu)}_2 \bar\Psi_\mu \gamma_5 \gamma^\lambda i\partial_\lambda \Psi_\nu$ &  $ q^{(\mu\nu)}_2 {P_\mu}^\upsilon{P_\nu}^\tau \bar\Psi_\upsilon \gamma_5 \gamma^\lambda i\partial_\lambda \Psi_\tau $ & $q^{(\mu\nu)}_2 {T_\mu}^\upsilon{T_\nu}^\tau \bar\Psi_\upsilon \gamma_5 \gamma^\lambda i\partial_\lambda \Psi_\tau$ & even \\
        \cline{2-5} & $\eta^{\mu\nu} f^\lambda_1 \bar\Psi_\mu i\partial_\lambda \Psi_\nu$ & $-\eta^{\mu\nu} f^\lambda_1 {P_\lambda}^\omega \bar\Psi_\mu i\partial_\omega \Psi_\nu$ & $- \eta^{\mu\nu} f^\lambda_1 {T_\lambda}^\omega \bar\Psi_\mu i\partial_\omega \Psi_\nu$& odd \\
        \cline{2-5} &$-i f^\lambda_2 \bar\Psi_\mu \Sigma^{\mu\nu} i\partial_\lambda \Psi_\nu$ & $if^\lambda_2 {P_\lambda}^\omega \bar\Psi_\mu \Sigma^{\mu\nu} i\partial_\lambda \Psi_\nu$ & $if^\lambda_2 {T_\lambda}^\omega \bar\Psi_\mu \Sigma^{\mu\nu} i\partial_\lambda \Psi_\nu$ & even \\
        \cline{2-5} & $\eta^{\mu\nu} g^\lambda_1 \gamma_5\bar\Psi_\mu i\partial_\lambda \Psi_\nu$ & $-\eta^{\mu\nu} f^\lambda_1 {P_\lambda}^\omega \bar\Psi_\mu \gamma_5 i\partial_\omega \Psi_\nu$ & $\eta^{\mu\nu} f^\lambda_1 {T_\lambda}^\omega \bar\Psi_\mu \gamma_5 i\partial_\omega \Psi_\nu$ & even \\
        \cline{2-5} &$-i g^\lambda_2 \bar\Psi_\mu \gamma_5 \Sigma^{\mu\nu} i\partial_\lambda \Psi_\nu$ & $-i g^\lambda_2 {P_\lambda}^\omega\bar\Psi_\mu \gamma_5 \Sigma^{\mu\nu} i\partial_\omega \Psi_\nu$ & $ig^\lambda_2 {T_\lambda}^\omega \bar\Psi_\mu \Sigma^{\mu\nu} i\partial_\lambda \Psi_\nu$ & odd \\
        \hline
    \end{tabular}
    \label{Lorentz-V.CPT}
\end{table}

A simple realization of its coefficients is: ${{C_s}^{(\mu\nu)}}_\lambda = \eta^{\mu\nu} a_\lambda$,  ${D_s}^{(\mu\nu)\lambda} = \eta^{\mu\nu} b^\lambda$ with $a^\lambda, b^\lambda$ constant vectors with mass dimension which are invariant under CPT. The second term has already been studied recently in the RS theory for $A=-1$, and has been associated with a Carroll-Field-Jackiw (CFJ) term which is a modification of standard electromagnetism~\cite{Gomes2023lorentz,Gomes2023non}. Other examples are ${{C_s}^{(\mu\nu)}}_\lambda = s_1^{\mu\nu} \bar{a}_\lambda$, ${{C_a}^{[\mu\nu]\lambda}} = \epsilon^{\mu\rho\nu\lambda} {a_1}_\rho$, ${D_s}^{\mu\nu\lambda} = s_2^{\mu\nu} \bar{b}^\lambda$, and ${D_a}^{\mu\nu\lambda} = \epsilon^{\mu\rho\nu\lambda}{a_2}_\rho$ (where $[s_1^{\mu\nu}] = [s_2^{\mu\nu}] = \text{mass}$ and axial vectors as well, while the other elements are dimensionless).

In summary, the extended RS Lagrangian that includes terms that violate Lorentz invariance has the form (for $k \leq 1$)
    \be
        {\cal L}_{\text{RS}}^{\text{Ex}} = \bar\Psi_\mu (\Lambda^{\mu\nu} - \tilde{\Lambda}^{\mu\nu}_0 - \tilde{\Lambda}^{\mu\nu}_1) \Psi_\nu \, .
    \ee
Another important property of RS theory is that the Lagrangian is invariant under the contact/point transformation, note that this statement is no longer valid with the inclusion of the LIV terms. 

Hadronic particles (mesons and baryons) are composed of quarks bound together by gluons. Any violation of Lorentz invariance and CPT symmetries in these composite particles must originate at the fundamental level—in the interactions between quarks and gluons. The ChPT provides a framework to connect the Lagrangian at the quark and gluon level with effective theories at the hadronic level, and an extension of this theory incorporating LIV has been proposed~\cite{Altschul2019lorentz}. However, describing spin-$\frac{3}{2}$ fermions requires the introduction of the particular RS Lagrangian~\cite{Rarita1941theory}. As a result, there is no complete connection for extending the LIV framework in this context. Using the operator classification presented here, one could characterize these operators in the context of ChPT to establish bounds on Lorentz violation in the hadronic regime.

In fact, the tensor coefficients presented here are constructed to preserve the hermiticity of the general RS Lagrangian density. These coefficients can be interpreted as arising from explicit Lorentz violation or from an unknown underlying mechanism. Once the philosophy of the SME, for example, is understood--namely, that it represents the most general observer-independent field theory with Lorentz violation--any Lagrangian involving known low-energy fields will ultimately be encompassed within the SME + this framework, which we will call a ``full SME''.
\section{Conclusions and outlook} \label{Conclusions}
In this paper, we present a new application of the chiral fermion formalism, we discuss the discrete symmetries for the general Rarita-Schwinger spin-$\frac{3}{2}$ field. We note that this theory features terms that are CP and T-odd, breaking the CP and T symmetries in some formulations, but still preserving CPT by virtue of the CPT theorem. We have also shown that treating the free RS field as a vector-spinor allows us to clearly formulate its transformation under the full Lorentz group, which includes the discrete symmetries.

Although our goals are partially to present a comprehensive and self-consistent review, we also discuss the advantages of using the two-component spinors as the basic building blocks for formulating CPT-violating QFTs.   
With this formalism, we have successfully classified RS theory + terms violating Lorentz and/or CPT symmetry while maintaining the linearity of the Lagrangian up to the first order in derivatives. This echoes the structure commonly seen in theories involving spin-$\frac{1}{2}$ fermions, such as in the extensions of QED. Our approach provides a framework that is consistent with effective field theory, allowing for the controlled introduction of LIV terms.

While we do not specify an exact mechanism for the origin of Lorentz violation, being inspired by the spontaneous Lorentz symmetry breaking, the formulation remains compatible with effective theories that often arise from high-energy physics scenarios. These may include string theory or other model frameworks beyond the standard one where Lorentz symmetry is not fundamental but emerges at low energies. This formalism is particularly useful for future studies aiming to explore Lorentz violation and/or CPT in more complex scenarios, such as supersymmetric extensions of gravity or higher spin field theories. Furthermore, this work opens the door to the investigation of the physical consequences of such violations in astrophysical phenomena, cosmology, and possible experimental searches in spin-$\frac{3}{2}$ baryons, for example, thus contributing to a deeper understanding of fundamental symmetries in nature.

\acknowledgments
JLDC acknowledges support from SNI (CONAHCYT) and VIEP (BUAP), APL wants to thank FCFM-BUAP for the warm and enjoyable hospitality during his sabbatical leave and where part of this work was done. The work of APL has been partially supported by CONAHCYT, Mexico, under grant 237004. IPC acknowledges financial support from CONAHCYT, Mexico, and thanks Jonathan Reyes Pérez for his valuable discussions during this work.

\appendix
\section{Dimensional reduction for vector-spinor fields: Gram-Schmidt Process and spin projectors} \label{D.red}
The Gram-Schmidt process is an algorithm for constructing another set of orthonormal vectors generating the same vector subspace from a set of linearly independent vectors forming a subspace. Calling the projection of $v$ on $u$ as $P_u v$, we have
    \be
        P_u v := \hat{u} \langle v, \hat{u} \rangle = u \frac{\langle v , u \rangle}{\langle u , u \rangle}\, .
    \ee
where $\hat{u} = \frac{u}{|u|}$, with $|u|$ the modulus of the vector $u$, and $\langle u , v \rangle$ is an inner product. Given the set of vectors $v_1, \ldots, v_n$, we can construct the set of orthogonal vectors $u_1, \ldots, u_n$ as follows,
    \be
        u_1 = v_1 \, ,
        u_2 = \left(1 - P_{u_1}\right) v_2 \, ,
        u_3 = \left(1 - P_{u_1} - P_{u_2}\right) v_3 \, , \dots, 
        u_n = \left(1 - \sum_{j=1}^{n-1} P_{u_j}\right) v_n \,.
    \ee
Following these ideas, we can find the components of the $\frac{3}{2}$-spinor of the Rarita-Schwinger field $\Psi^\mu$ in $d$-dimensional formulation ($\mu = 0,1, \dots, d-1$), doing the following correspondences
    \be
        v_1 \rightarrow p^\mu I_d \, , \quad
        v_2 \rightarrow \gamma^\mu  \, , \quad 
        v_3 \rightarrow \Psi^\mu \, .
    \ee
with $I_d$ identity matrix $d \times d$ (which we omitted in the direct calculations, but is present) and defining the following correspondence between ``vectors'':
    \be
        \hat{u}_1 \,\rightarrow\, \hat{p}^\mu = \frac{p^\mu}{|p|} \, , 
        u_2 \,\rightarrow\, \gamma^\mu - \frac{ \slashed{p} p^\mu}{|p|^2} \, ,
    \ee
where we have
    \be
        |u_2|^2 = (u_2)^\mu (u_2)_\mu \rightarrow \left( \gamma^\mu - \frac{(\gamma^\nu p_\nu) p^\mu}{|p|^2} \right) \left( \gamma_\mu - \frac{(\gamma^\rho p_\rho) p_\mu}{|p|^2} \right) = \gamma^\mu\gamma_\mu - \frac{(\gamma^\rho p_\rho) (\gamma^\mu p_\mu)}{|p|^2} = (d-1) I_d  \, .
    \ee
Here we have used $\slashed{p}\slashed{p} = |p|^2 I_d$. Then $(u_2)^\mu \rightarrow \frac{1}{\sqrt{d-1}} \left( \gamma^\mu - \frac{(\gamma^\nu p_\nu) p^\mu}{|p|^2} \right) := \Lambda^\mu$, where we have introduced the ``vector'' $\Lambda^\mu$. 
Now, we define the projectors
    \be
        (P_{22}^{1/2})^{\mu\nu} = \frac{p^\mu p^\nu}{|p|^2} \quad , \quad 
        (P_{11}^{1/2})^{\mu\nu} = \Lambda^\mu \Lambda_\nu \, ,
    \ee
with $\Lambda^\mu\Lambda^\nu = \frac{1}{\sqrt{d-1}} \left( \gamma^\mu - \frac{ \slashed{p} p^\mu}{|p|^2} \right) \frac{1}{\sqrt{d-1}} \left( \gamma^\nu - \frac{ \slashed{p} p^\nu}{|p|^2} \right) = \frac{1}{d-1} \left( \gamma^\mu \gamma^\nu + \frac{1}{|p|^2} \slashed{p} (\gamma^\mu p^\nu - \gamma^\nu p^\mu) - \frac{p^\mu p^\nu}{|p|^2} \right) $.
Taking now, $(v_3)^\mu \rightarrow \Psi^\mu$, the Rarita-Schwinger vector-spinor for the gravitino can be written in terms of the projectors as:
    \be
        \Psi_{3/2}^\mu = (P^{3/2})^{\mu\nu} \Psi_\nu = \left[ \eta^{\mu\nu} - (P_{22}^{1/2})^{\mu\nu} - (P_{11}^{1/2})^{\mu\nu} \right] \Psi_\nu \, .
    \ee
With spin projector (with $|p|^2 = p^2$)
    \be
        (P^{3/2})^{\mu\nu} = \eta^{\mu\nu} - \frac{1}{(d-1)}  \gamma^\mu \gamma^\nu  -  \frac{1}{(d-1) p^2}\slashed{p} \left( \gamma^\mu p^\nu - p^\mu \gamma^\nu \right) - \frac{(d-2)}{(d-1)}\frac{p^\mu p^\nu}{p^2}  \, .
    \ee
Note that it is convenient to rewrite
    \be
    \begin{aligned}
       (P_{11}^{1/2})^{\mu\nu} &= \frac{1}{(d-1) p^2} \left( \gamma^\mu \slashed{p}^2 \gamma^\nu + \slashed{p} \left( \gamma^\mu p^\nu - p^\mu \gamma^\nu \right) - p^\mu p^\nu \right) \\
       &= -\frac{1}{(d-1) p^2} \left( \slashed{p} \gamma^\mu \slashed{p} \gamma^\nu - 2 p^\mu \slashed{p} \gamma^\nu - \slashed{p} \left( \gamma^\mu p^\nu - p^\mu \gamma^\nu \right) + p^\mu p^\nu \right) \\
       &= \frac{1}{(d-1)p^2} i\left(p^\mu - \slashed{p}\gamma^\mu\right) i\left(p^\nu - \slashed{p}\gamma^\nu \right) = i\omega^\mu i\omega^\nu \, \quad \text{with} \quad \omega^\mu := \frac{1}{\sqrt{d-1} p} \left(p^\mu - \slashed{p}\gamma^\mu\right) \, .
    \end{aligned}
    \ee
We can construct the following operators
    \be
        (P_{12}^{1/2})^{\mu\nu} = \omega^\mu \frac{p^\nu}{p} = \frac{1}{\sqrt{d-1}p^2} \left(p^\mu p^\nu - \slashed{p}\gamma^\mu p^\nu \right) \quad , \quad
        (P_{21}^{1/2})^{\mu\nu} = - \frac{p^\mu}{p} \omega^\nu = \frac{1}{\sqrt{d-1}p^2} \left(\slashed{p} p^\mu \gamma^\nu  - p^\mu p^\nu \right)  \, ,
    \ee
These projectors are mentioned in the literature~\cite{Van1981supergravity}. 
An important property of these projectors (for example in the construction of the general operator $\Lambda^{\mu\nu}$) is
    \be \label{Special}
        \left( P^I_{ij} \right)_{\mu \nu} \left( P^J_{kl} \right)^{\nu \rho} = \delta^{IJ} \delta_{jk} {\left( P^I_{il} \right)_{\mu}}^{\rho}, \quad \text{with} \quad I, J \in \left\{ \frac{1}{2}, \frac{3}{2} \right\}, \ i, j, k, l \in \{1, 2\}.
    \ee
\bibliography{ref}

\end{document}